\documentclass{aa}  
\usepackage{graphicx}
\begin{document}
   \title{Deep Extragalactic VLBI-Optical Survey (DEVOS)}

   \subtitle{II. Efficient VLBI detection of SDSS quasars\thanks{Table~\ref{image-par} and Fig.~\ref{images} are only available in electronic form at {\tt http://www.edpsciences.org}}}

   \titlerunning{Deep Extragalactic VLBI-Optical Survey (DEVOS) II.}

   \author{S. Frey\inst{1,2} 
           \and
           L.I. Gurvits\inst{3}
           \and
           Z. Paragi\inst{3,2}
           \and
           L. Mosoni\inst{4,5}
           \and
           M.A. Garrett\inst{6}
           \and
           S.T. Garrington\inst{7} 
          }

   \offprints{S. Frey}

   \authorrunning{S. Frey et al.}

   \institute{F\"OMI Satellite Geodetic Observatory, P.O. Box 585, H-1592 Budapest, Hungary\\
              \email{frey@sgo.fomi.hu} 
              \and
              MTA Research Group for Physical Geodesy and Geodynamics, P.O. Box 91, H-1521 Budapest, Hungary 
              \and
              Joint Institute for VLBI in Europe, Postbus 2, 7990 AA Dwingeloo, The Netherlands
              \and
              Konkoly Observatory of the Hungarian Academy of Sciences, P.O. Box 67, H-1525, Budapest, Hungary
              \and
              Max-Planck-Institut f\"ur Astronomie, K\"onigstuhl 17, D-69117 Heidelberg, Germany
              \and
              Netherlands Foundation for Research in Astronomy (ASTRON), Postbus 2, 7990 AA Dwingeloo, The Netherlands
              \and
              University of Manchester, Jodrell Bank Observatory, Macclesfield, Cheshire SK11 9DL, UK
              }

   \date{Received 20 September 2007 / Accepted 2 November 2007}

 
  \abstract
{The Deep Extragalactic VLBI-Optical Survey (DEVOS) aims at
constructing a large sample of compact radio sources up to two orders of 
magnitude fainter than those studied in other Very Long Baseline
Interferometry (VLBI) surveys. Optical identification of the objects is ensured
by selecting them from the Sloan Digital Sky Survey (SDSS) list.}
{While continuing to build up the DEVOS data base, we investigated how the VLBI detection rate could be enhanced by refining the initial selection criteria introduced in the first paper of this series.}
{We observed 26 sources in two adjacent, slightly overlapping $2\degr$-radius fields with the European VLBI Network (EVN) at 5~GHz frequency on 2 March 2007.
The phase-reference calibrator quasars were J1616+3621 and J1623+3909. 
The objects selected were unresolved both in the Faint Images of the Radio Sky at Twenty-centimeters (FIRST) survey catalogue 
and the SDSS Data Release 4.}
{We present images of milli-arcsecond (mas) scale radio structures and accurate coordinates of 24 extragalactic sources. 
Most of them have never been imaged with VLBI. Twenty-two compact radio sources (85\% of our initial sample) are considered 
as VLBI detections of the corresponding optical quasars in SDSS. 
We found an efficient way to identify quasars as potential VLBI targets with mas-scale compact radio stucture at $>1$~mJy level, based only on the FIRST and SDSS catalogue data by applying simple selection criteria.}
   {}


   \maketitle
%

\section{Introduction}

Distant extragalactic radio sources detected with Very Long Baseline Interferometry (VLBI) are typically active galactic nuclei (AGNs): the rest-frame brightness temperature of a 1-mJy source radiating from a region subtending a few milli-arcseconds (mas) at 5~GHz is at least $10^6 - 10^7$~K, often much higher.
The aim of the Deep Extragalactic VLBI-Optical Survey (DEVOS; Mosoni et al. \cite{moso06}, Paper~I hereafter) is to build up a large sample of optically identified AGNs imaged with VLBI at mas angular resolution. Many of the potential astrophysical and cosmological utilisations of such a sample require the optical identification and spectroscopic (or photometric) redshift of the sources. These data can primarily be obtained from large optical sky surveys. In particular, the latest release\footnote{Data Release 6 (DR6), \tt{http://www.sdss.org/dr6}} of the Sloan Digital Sky Survey (SDSS; Adelman-McCarthy et al. \cite{adel07}) contains imaging data for over 20\,000, and spectroscopic data for more than 3000 unresolved radio sources from the Faint Images of the Radio 
Sky at Twenty-centimeters (FIRST) survey catalogue (White et al. \cite{whit97})\footnote{\tt http://sundog.stsci.edu} with integrated flux density of at least 20~mJy. These sources could be targeted with phase-referenced radio interferometric observations by selecting nearby bright and compact calibrator sources from e.g. the VLBA Calibrator Survey (VCS)\footnote{\tt http://www.vlba.nrao.edu/astro/calib}. To indicate the potential of a FIRST- and SDSS-based VLBI imaging survey, we note that optical counterparts of at least 30\% of all FIRST sources are expected to be found in SDSS (Ivezi\'c et al. \cite{ivez02}). A recent sensitive single-baseline Arecibo--Effelsberg VLBI survey of $\sim1000$ FIRST sources indicated that about one third of {\em all} sources had mJy-level compact stucture, irrespective of the total flux density (Porcas et al. \cite{porc04}).

The technical feasibility of DEVOS has recently been demonstrated in a pilot experiment (Paper~I). However, at present reaching the ambitious goal of increasing the total number of VLBI-imaged extragalactic sources to $\sim10^4$ is very demanding in terms of observing resources. This led us to somewhat modify the sample definition criteria used in Paper I in order to increase the throughput of the survey. We studied two adjacent celestial fields, each of $2\degr$ radius, to search for mas-scale structures of radio-emitting quasars detected with the SDSS. Here we describe the target selection (Sect.~\ref{sect2}), the phase-referenced VLBI observations (Sect.~\ref{sect3}) and the data reduction procedure (Sect.~\ref{sect4}). The results are presented in Sect.~\ref{sect5}. The VLBI images of the target sources are available in the online version of this paper. Implications of our results are discussed in Sect.~\ref{sect6}.


\section{Sample selection}
\label{sect2}

The survey approach for the DEVOS pilot project (Paper I) was to select unresolved sources from the FIRST survey (White et al. \cite{whit97}) with integrated flux density $S_{1.4}>30$~mJy in the vicinity (within $2^{\circ}$) of a selected calibrator source. Then 5-GHz phase-referencing observations with the UK Multi-Element Radio Linked Interferometer Network (MERLIN) were performed to pinpoint potential compact sources suitable as VLBI imaging targets. In the end, 19 sources (40\% of the original FIRST sample) could be detected with a global network of VLBI antennas at 5~GHz. These DEVOS pilot observations showed, among others things, that all the 6 FIRST sources that appeared stellar in SDSS imaging (i.e. optical quasars) could be detected and imaged with VLBI. 
In the case of these objects, the offsets between the VLBI and SDSS positions were between 30 and 90~mas. Independent comparisons of accurate radio interferometric source positions with the optical coordinates of their SDSS counterparts (Frey et al. \cite{frey06a,frey06b}; Lambert et al. \cite{lamb06}) also indicate that SDSS coordinates of radio-emitting quasars are accurate enough to be used as a priori values for phase-referenced VLBI observations.

Building on this experience, we modified the sample selection criteria applied in Paper I, concentrating on SDSS {\em quasars} only. The reason is to find a way to speed up the construction of the radio--optical data base by means of $(i)$ reducing the radio observing resources required, in particular eliminating the need of ``filtering'' MERLIN observations, and $(ii)$ considerably improving the detection rate of the VLBI observations. These improvements are achieved by tightening the original selection criteria. This comes to an expense of losing some of the generality of the DEVOS approach and avoiding potentially interesting individual objects or classes. We stress, however, that the survey can be supplemented with the sources missed by these more stringent criteria at a later stage.  

With this compromise, the objectives of many potential scientific applications of DEVOS outlined in Paper I can still be addressed. In particular, the morphology of low-luminosity quasars could be compared with that of the high-luminosity ones; cosmological tests could be conducted with substantially larger samples; precise astrometric positions could be used to directly link the radio and future optical reference frames; the sources detected could serve as phase-reference calibrators --~either in the traditional ``nodding'' style or in the in-beam mode~-- for other VLBI projects.

The target extragalactic radio sources for the observations described here
were chosen from the FIRST survey data base obtained with the
US National Radio Astronomy Observatory (NRAO) Very Large Array (VLA)
at 1.4~GHz (White et al. \cite{whit97}). Their optical identification is ensured by cross-comparing the list with the 4th Data Release (DR4)\footnote{\tt http://www.sdss.org/dr4} of the SDSS (Adelman-McCarthy et al. \cite{adel06})
-- the actual release at the time of preparing for the observations. The selection criteria were as follows:

\begin{enumerate}
  \item the integrated flux density at 1.4~GHz is $S_{1.4} > 20$~mJy;
  \item the angular size of the source is $\theta < 5\arcsec$
  (i.e. unresolved with the VLA in the FIRST);
  \item the angular separation of the target source from the
  phase-reference calibrator source selected is less than $2\degr$;
  \item the source has an unresolved (stellar) optical counterpart in SDSS DR4 within $1\farcs5$ from the FIRST position.
\end{enumerate}

Note that the minimum flux density is lower here than that applied in Paper I (30~mJy). Very few Galactic radio stars would satisfy our selection criteria, thus the extragalactic nature of the selected objects is practically ensured and could be verified with SDSS spectral data.   

Radio interferometric images of compact sources that are up to 2 orders of magnitude weaker than currently available in other 
large-scale VLBI surveys can be made with the technique of phase-referencing (e.g. Lestrade et al. \cite{lest90};
Beasley \& Conway \cite{beas95}). It is based on extending the coherence time by using regularly
interleaving observations of a nearby bright and compact calibrator source.
Here we select suitable calibrators first, then look for the potential target objects that are found in their close vicinity in the sky.
Thus single calibrators can serve as phase-reference sources
for several targets. This method was introduced by Garrett \& Garrington (\cite{garr98}) and
Garrington et al. (\cite{garr99}).

We choose two bright compact radio sources (J1616+3621 and J1623+3909) as phase-reference calibrators. These are known VLBI sources (Beasley et al. \cite{beas02}; Helmboldt et al. \cite{helm07}) and have a relatively large number (13 and 16) of matching FIRST/SDSS quasars in their neighbourhood, according to the selection criteria above. The equatorial coordinates of the calibrators are determined in the International Celestial Reference Frame (ICRF) with mas or sub-mas accuracy (Petrov \cite{petr07}), allowing us to determine the high-accuracy ICRF positions of the target sources through relative VLBI astrometry. There is a small overlap (3 common sources) between the two fields (Field A around J1616+3621 and Field B around J1623+3909). By comparing the images and positions derived independently for these common targets, we can assess the reliability of our survey method. 

\begin{figure}
 \centering
  \includegraphics[width=80mm]{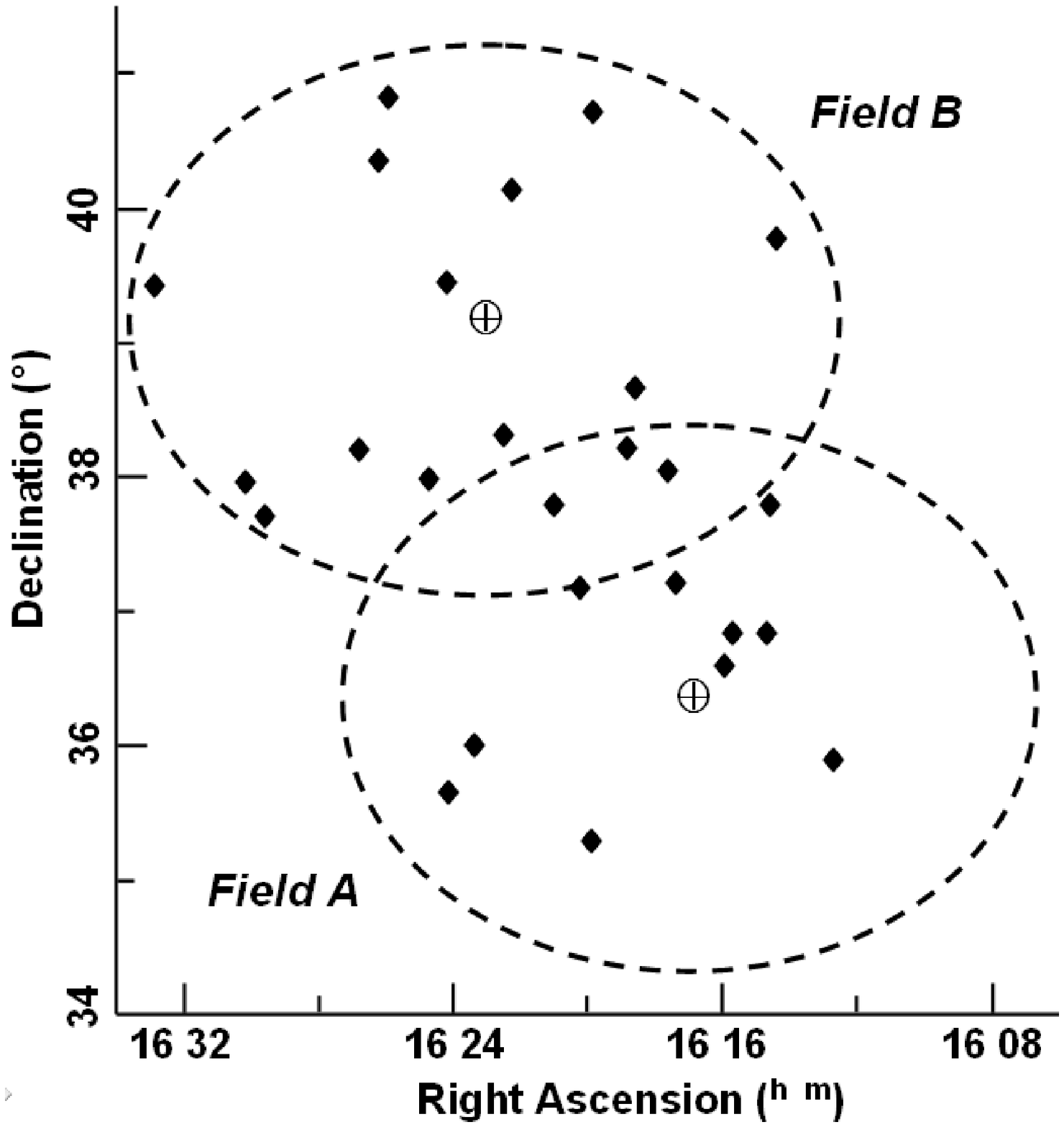}
  \caption{Sketch of the celestial positions of the two phase-reference sources ($\oplus$) and the 26 programme sources. The borders of Fields~A and B projected onto the right ascension--declination plane are indicated with dashed lines.}
\label{skyplot}
\end{figure}

The list of target sources is given in Table~\ref{fieldA} for both fields. The source names were derived from the VLBI coordinates wherever available, otherwise the SDSS coordinates were used. For reference, we list the SDSS spectroscopic redshift (if available), the SDSS $r$ magnitude and the FIRST integrated flux density at 1.4~GHz. We also give the 5-GHz total flux densities from the GB6 catalogue (Gregory et al. \cite{greg96}). Together with the FIRST values, these are useful to infer the radio spectral slope of the sources. However, due to their detection limit of $\sim18$~mJy, our weaker targets are not found in the GB6 catalogue. The approximate positions of the reference and target sources observed are shown in Fig.~\ref{skyplot}.

\begin{table*}
 \caption{The 13 and 16 sources targeted around the phase-reference calibrators J1616+3621 (Field~A, {\it top}) and J1623+3909 (Field~B, {\it bottom})}
 \label{fieldA}
 \centering
 \begin{tabular}{@{}lllrrrrrrr}
  \hline\hline
  Source name & \multicolumn{2}{c}{Equatorial coordinates (J2000)} & $z$ & SDSS & FIRST & GB6 &
  \multicolumn{2}{c}{VLBI 5~GHz} & offset from \\
              & R.A. & Dec.                                        &     & $r$  & int   & int & 
                    peak & int & SDSS position\\
  & h\hspace{3mm}m\hspace{3mm}s & $\degr$\hspace{4mm}$\arcmin$\hspace{3mm}$\arcsec$ & & & mJy & mJy & mJy/bm & mJy & mas \\
  \hline
J161252.4+355234  & 16 12 52.42647 & 35 52 34.3661 &      &  22.0 &  21.6 &    -- &   3.7 &   8.0 &  23 \\
J161446.9+374607  & 16 14 46.95806 & 37 46 07.2795 & 1.53 &  16.6 &  51.1 &    36 &  76.9 &  85.2 & 108 \\ 
J161453.1+364908  & 16 14 53.10423 & 36 49 08.5616 &      &  22.8 &  85.8 &    35 &   6.8 &  28.6 & 347 \\ 
J161550.9+364827  & 16 15 50.93939 & 36 48 27.3383 &      &  21.1 &  24.3 &    22 &   4.8 &   8.7 &  64 \\
J161606.0+363614  & 16 16 06.01644 & 36 36 14.4327 & 1.36 &  19.9 &  24.9 &    22 &  14.0 &  17.9 &  33 \\
J161734.3+371154  & 16 17 34.34556 & 37 11 54.9837 &      &  21.0 &  32.6 &    -- &   4.1 &  12.9 &  94 \\
J161748.4+380141* & {\it 16 17 48.41146} & {\it 38 01 41.8062} & 1.61 &  19.2 &  21.2 &    60 &  61.4 &  70.4 &  53 \\
J161900.7+381139* & {\it 16 19 00.73053} & {\it 38 11 39.3010} &      &  21.4 &  44.9 &    -- &   3.8 &   6.5 &  51 \\
J162004.7+351554  & 16 20 04.73836 & 35 15 54.4634 & 2.94 &  18.6 &  50.0 &    38 &  27.5 &  38.5 & 174 \\
J162027.0+371026  & 16 20 27.03237 & 37 10 26.8544 &      &  20.4 &  77.2 &    21 &   1.3 &   1.4 & 156 \\
J162111.2+374604* & {\it 16 21 11.27611} & {\it 37 46 04.8423} & 1.27 &  18.8 & 643.1 &   201 &  13.5 &  46.7 &  80 \\
J162330.5+355933  & 16 23 30.53332 & 35 59 33.1456 & 0.87 &  18.3 & 266.8 &   144 &  83.5 & 131.2 & 110 \\
J162419.9+353845  & 16 24 19.95778 & 35 38 45.2701 & 1.77 &  18.2 &  28.8 &    22 &  16.3 &  20.0 &  47 \\
  \hline
J161432.1+394445  & 16 14 32.19838 & 39 44 45.0770 &      &  20.6 & 316.1 &   134 &   0.9 &   1.6 & 675 \\ 
J161748.4+380141* & 16 17 48.41133 & 38 01 41.8048 & 1.61 &  19.2 &  21.2 &    60 &  63.2 &  70.1 &  54 \\ 
J161845.2+383807  & 16 18 45.28153 & 38 38 07.7917 &      &  19.6 &  90.7 &    46 &   4.2 &   8.1 &  48 \\ 
J161900.7+381139* & 16 19 00.73060 & 38 11 39.3000 &      &  21.4 &  44.9 &    -- &   3.9 &   5.6 &  50 \\ 
J162000.4+404319  & 16 20 00.47630 & 40 43 19.1665 &      &  22.4 & 103.9 &    27 &   4.4 &  11.9 & 107 \\ 
J162111.2+374604* & 16 21 11.27601 & 37 46 04.8406 & 1.27 &  18.8 & 643.1 &   201 &  12.9 &  20.6 &  82 \\ 
J162229.3+400643  & 16 22 29.31193 & 40 06 43.5915 & 0.69 &  18.6 &  32.1 &    58 &  75.1 &  83.5 &  88 \\ 
J162240.7+381637  & 16 22 40.72642 & 38 16 37.3210 &      &  22.5 &  73.5 &    83 &  35.8 &  39.8 &  48 \\ 
J162422.0+392440  & 16 24 22.00133 & 39 24 40.9144 & 1.12 &  17.9 & 144.5 &   128 &  87.3 &  97.6 &  59 \\ 
J162453.4+375806  & 16 24 53.47942 & 37 58 06.6459 & 3.38 &  18.5 &  56.4 &    22 &   8.6 &  23.6 &  11 \\ 
J162605.1+404806  & 16 26 05.15018 & 40 48 06.6508 &      &  22.0 & 125.6 &    -- &   1.7 &   5.2 & 773 \\ 
J162624.9+401945  & 16 26 24.95879 & 40 19 45.4248 &      &  19.2 & 104.6 &    39 &   0.7 &   0.8 &  66 \\ 
J162656.5+381000  & 16 26 56.52929 & 38 10 00.9853 &      &  20.3 &  54.2 &    29 &  20.1 &  24.8 & 123 \\ 
J162945.0+374003  & 16 29 45.0826  & 37 40 03.409  &      &  20.9 &  33.0 &    -- & $<0.6$ &   --  & -- \\ 
J163020.7+375656  & 16 30 20.7733  & 37 56 56.412  & 0.39 &  17.9 &  21.7 &    -- & $<0.7$ &   --  & -- \\ 
J163302.1+392427  & 16 33 02.10492 & 39 24 27.3907 & 1.02 &  16.6 &  54.4 &    35 &  13.1 &  19.6 &  53 \\ 
  \hline
  \end{tabular}
\\
Notes: Col.~1 -- source name derived from the VLBI coordinates (if detected) or the SDSS coordinates;
sources marked with * are observed in both fields; for these common sources the Field~B coordinates are more accurate, 
but we also list the Field~A coordinates typed in italics for comparison;   
Col.~2 -- J2000 right ascension (h~m~s);
Col.~3 -- J2000 declination ($\degr$~$\arcmin$~$\arcsec$);
Col.~4 -- spectroscopic redshift;
Col.~5 -- SDSS $r$ magnitude;
Col.~6 -- FIRST integrated flux density at 1.4~GHz (mJy); 
Col.~7 -- GB6 total flux density at 5~GHz (mJy);
Col.~8 -- VLBI peak brightness at 5~GHz (mJy/beam), an upper limit is given for non-detections;
Col.~9 -- VLBI integrated flux density at 5~GHz (mJy);
Col.~10 -- VLBI peak--SDSS position difference (mas) 
 \end{table*}


\section{Observations}
\label{sect3}

The 5-GHz VLBI observations involving 8 antennas of the European VLBI Network (EVN)
took place on 2 March 2007. The total observing time was 6 hours.
The participating radio telescopes were Effelsberg (Germany), the phased array of the 14-element 
Westerbork Synthesis Radio Telescope (The Netherlands), Jodrell Bank Mk2 
(UK), Noto (Italy), Toru\'{n} (Poland), Onsala (Sweden), Sheshan and Nanshan
(P.R. China).
The recording data rate was 1 Gbit s$^{-1}$. Eight 
intermediate frequency channels (IFs) were used in both left and right circular 
polarisation. The total bandwidth was 128~MHz in both polarisations.

\begin{figure}[b]
 \centering
  \includegraphics[width=85mm,bb= 75 125 580 622,angle=270,clip=]{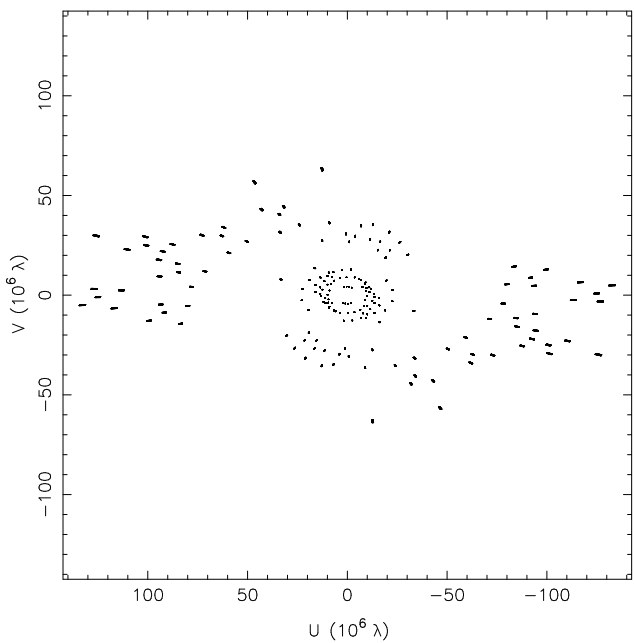}
  \caption{The $(u,v)$-coverage of one of the
target sources (J161252.4+355234) in our 5-GHz EVN observations.}
\label{uvpl}
\end{figure}

We observed a calibrator and two programme sources in $\sim5$-min phase-referencing cycles which included the antenna slewing and the time spent on the calibrator source.
Each programme source was observed in 4 scans of $\sim2$~min duration. These scans were distributed over the whole observing period to obtain a better ($u,v$)-coverage of which an example is shown in Fig.~\ref{uvpl}. 
The typical restoring beam using natural weighting is $\sim4.1$~mas~$\times$~1.4~mas in a position angle 
(PA) of $-5\degr$.
The theoretical image thermal noise ($1\sigma$) was estimated as $\sim70$~$\mu$Jy/beam.
The correlation took place at the EVN Data Processor at the Joint 
Institute for VLBI in Europe (JIVE) in Dwingeloo (The Netherlands).
The integration time was 2 s, the undistorted field of view was limited to $5\arcsec$ by bandwidth-smearing.


\begin{figure}
\centerline{
  \includegraphics[width=80mm,bb= 70 174 505 621,angle=270,clip= ]{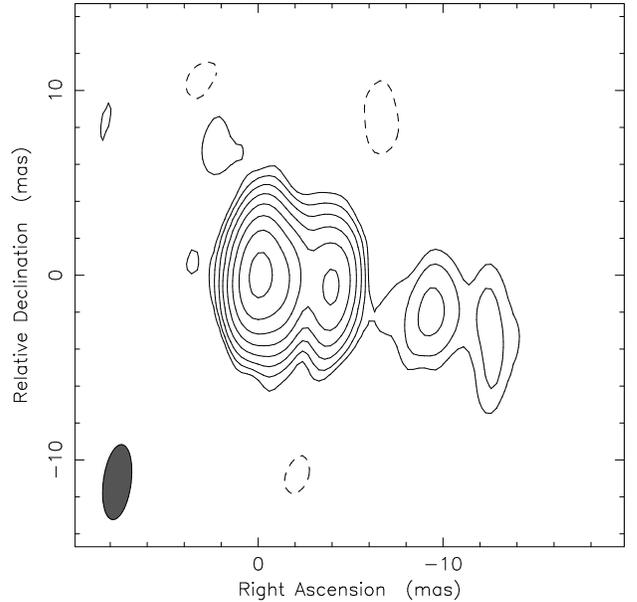}}
  \caption{Naturally weighted 5-GHz VLBI image of the reference source in Field~A 
(J1616+3621). The lowest contours are drawn at $\pm0.5$~mJy/beam. The
positive contour levels increase by a factor of 2. The peak brightness is
81.4~mJy/beam. The restoring beam is $4.1$~mas~$\times$~$1.5$~mas at
PA=$-7\fdg4$. The coordinates are relative to the brightness peak. The
Gaussian restoring beam is indicated with an ellipse in the lower-left corner.}
\label{a-cal}
\end{figure}

\begin{figure}
\centerline{
  \includegraphics[width=80mm,bb= 70 174 505 621,angle=270,clip= ]{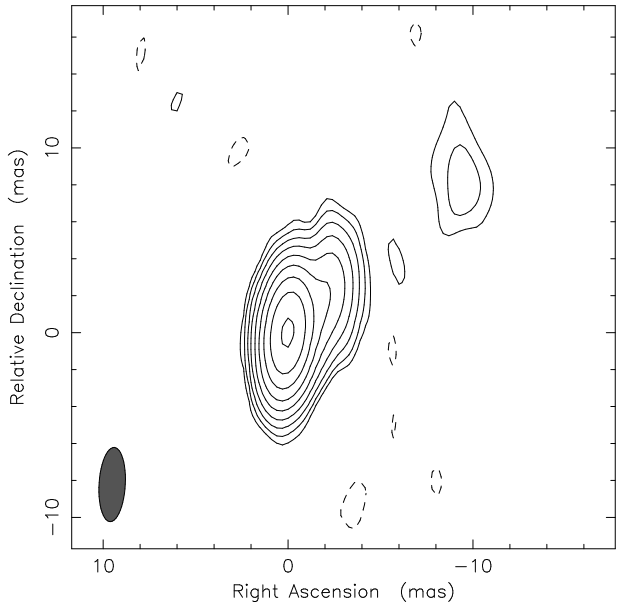}}
  \caption{Naturally weighted 5-GHz VLBI image of the reference source in Field~B 
(J1623+3909). The lowest contours are drawn at $\pm0.3$~mJy/beam. The
positive contour levels increase by a factor of 2. The peak brightness is
85.4~mJy/beam. The restoring beam is $4.0$~mas~$\times$~$1.4$~mas at
PA=$-3\fdg3$.}
\label{b-cal}
\end{figure}

\section{Data processing}
\label{sect4}

The NRAO Astronomical Image Processing System (AIPS) was used for the data calibration (e.g. Diamond \cite{diam95}). 
Visibility amplitudes were calibrated using system temperatures measured regularly at the antennas during the observations. 
Fringe-fitting was done separately for each phase-reference calibrator (J1616+3621, J1623+3909) and other secondary calibrator or fringe-finder sources (J1613+3412, J1642+3948). The calibrated visibility data were exported to the Caltech Difmap package (Shepherd et al. \cite{shep94}) for imaging. The conventional hybrid mapping procedure involving several iterations of CLEANing and phase (then amplitude) self-calibration resulted in the images shown in Figs.~\ref{a-cal} and \ref{b-cal} for the two phase-reference calibrators. Overall antenna gain correction factors were determined and applied to the visibility amplitudes of the target sources in AIPS. In any case, these corrections were not larger than 8\%. 

The fringe-fitting was repeated in AIPS, taking the calibrator clean component models into account.
The residual amplitude and phase corrections resulted from the non-pointlike structure of the phase-reference calibrators were considered this way. 
The phase, delay and delay-rate solutions obtained were interpolated and applied to the target source data. 
The visibility data of the programme sources, unaveraged in time and frequency, were also exported to Difmap for imaging.
Consistently with our analysis in Paper I, for the sources with integrated flux density less than 15~mJy, the phase-referenced images displayed in Fig.~\ref{images} are the results of a CLEANing process only. For the brighter sources, phase self-calibration with a solution interval of 30~s was also performed using the CLEAN component model. Amplitude self-calibration was only done for the sources brighter than $\sim40$~mJy. In the images of Fig.~\ref{images}, the coordinates shown are relative to the brightness peak position. However, the technique of phase-referencing preserves the sky position of the target sources. The absolute coordinates of the brightness peaks are given in Table~\ref{fieldA}. These positions were located using the AIPS task MAXFIT. The image parameters are listed in Table~\ref{image-par}.

Independently from the phase-referenced solutions, direct fringe-fitting for 8 strong target sources with peak brightness $>20$~mJy/beam (Col.~8 in Table~\ref{fieldA}) was done in AIPS. This allowed us to check the images in order to qualitatively verify the source structures. The image parameters (peak brightness, noise level) could also be compared with those of the phase-referenced images.
The dynamic range of the latters may be affected by residual errors introduced by the switching between the target and reference sources. By comparing our phase-referenced images with those obtained after direct fringe-fitting, we estimated the coherencence loss. The differences in peak brightness were ranging from 4\% to 23\%, with an average value of 12\%. The $5\sigma$ upper limits for our two non-detections (J162945.0+374003 and J163020.7+375656) were increased by 25\% to account for the loss of coherence (Table~\ref{fieldA}).


\section{Results and discussion}
\label{sect5}

The 5-GHz VLBI images of 24 sources are displayed in Fig.~\ref{images}. 
Most of them show nearly unresolved mas-scale radio structures, although typical extended one-sided core-jet
sources are also seen. The sources detected can be grouped according to their peak brightness as follows:

\begin{itemize}
 \item 5 bright objects ($\sim60-90$~mJy/beam)
 \item 7 objects with $\sim10-35$~mJy/beam
 \item 9 objects with $\sim2-8$~mJy/beam
 \item 3 weak objects ($\sim1$~mJy/beam)
\end{itemize}

In comparison with the results obtained in the DEVOS pilot experiment (Paper I), not just the detection rate is more favourable here (85\% vs. 40\% of the initial FIRST-based sample), but also the ratio of the bright objects within the detected sample is considerably higher.

We note that only about a half of the sources --~those with $r\la20^{\rm{m}}$~-- in our present sample have measured spectroscopic redshifts in SDSS. 
This is in fact due to the spectroscopic target selection method applied for quasars in SDSS (Richards et al. \cite{rich02}).
The relation between the optical $r$ magnitude (taken from the SDSS DR6) and the 5-GHz flux density in the VLBI-detected components is shown in Fig.~\ref{mag-fluxd}. Sources from Paper~I that satisfy our present selection criteria are also included in the plot. Although our sample is rather small for a rigorous statistical evaluation of this relation, it can be seen that the optically brighter sources tend to have higher radio flux densities in their VLBI components. In particular, the mean 5-GHz flux density is 53~mJy for the 12 quasars with known redshift ($r<20^{\rm{m}}$), while for the remaining 15 quasars ($r>19^{\rm{m}}$) this value is 23~mJy. Most of our compact sources that show nearly unresolved VLBI structures with no extended features (5 out of 6) are drawn from the brighter sub-sample.

\begin{figure}
\setcounter{figure}{5}
\centering
  \includegraphics[width=65mm, angle=270]{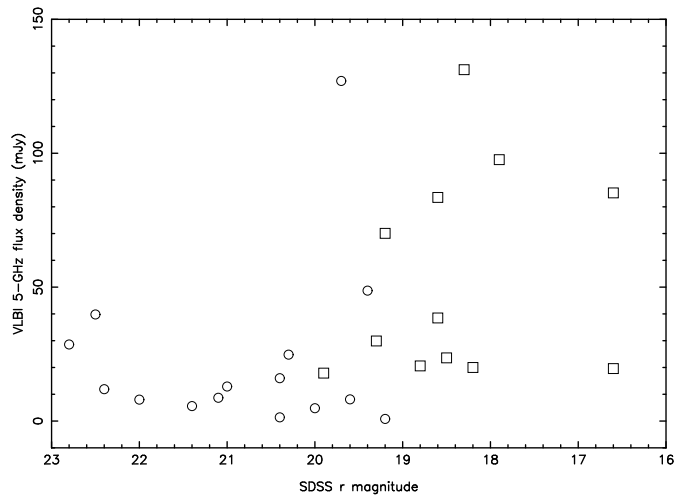}
  \caption{The 5-GHz VLBI flux density (mJy) versus the SDSS $r$ magnitude of the quasars detected with VLBI (this paper and Paper~I). Spectroscopic redshifts are available for the objects marked with squares.}
\label{mag-fluxd}
\end{figure}

The 5-GHz radio luminosity of the 12 VLBI-detected DEVOS quasars with known redshift (Paper~I and this work) is displayed in Fig.~\ref{lum-z}. The luminosity values are around $10^{26}$~W~Hz$^{-1}$. Indeed these are one or two orders of magnitude smaller than the typical luminosities in VLBI surveys of bright AGNs. To illustrate this, we also plotted in the same figure the VLBI Space Observatory Programme (VSOP) 5-GHz pre-launch VLBA survey data (VLBApls; Fomalont et al. \cite{foma00}) for comparison. Here we used the sum of the fitted model component flux densities to calculate luminosities of the sources with redshift values available (the overwhelming majority of the VLBApls sample). Note that the VLBApls sample also contains AGNs that are optically identified with galaxies. The quasar luminosities in our small sample of 12 sources are similar within a factor of $\sim14$, despite the broad range of redshifts ($0.69<z<3.38$) covered. Qualitatively, there is no obvious difference between the compact mas-scale radio structures of our low and high redshift quasars.

\begin{figure}
 \centering
  \includegraphics[width=65mm, angle=270]{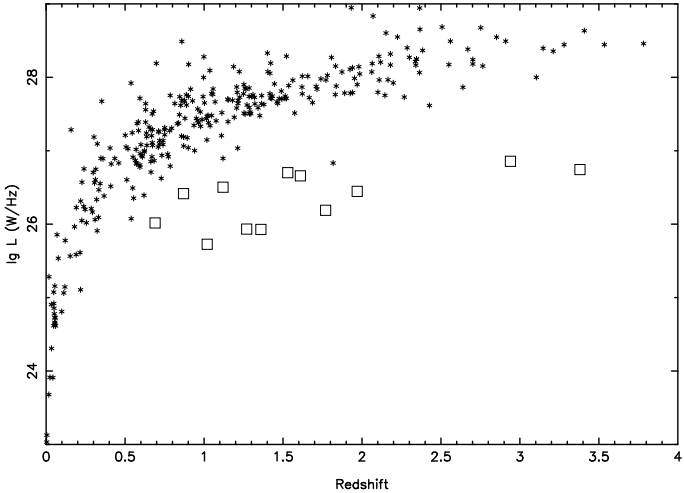}
  \caption{The integrated 5-GHz radio luminosity (W~Hz$^{-1}$) of the VLBI components as a function of the source redshift for the DEVOS sources with known redshift (squares) and the generally brighter radio-loud AGNs from Fomalont et al. (\cite{foma00}) (asterisks). }
\label{lum-z}
\end{figure}

\subsection{Astrometric accuracy}

The ICRF positional accuracy for the target sources depends on the phase-reference calibrator source position accuracy, the target--calibrator angular separation, the angular resolution of the interferometer array and the signal-to-noise ratio (see also the discussion in Paper I). We conservatively assume a relatively weak source (dirty image peak/rms ratio 5:1) at the edge of the field at $2\degr$ separation from the calibrator. Since a priori calibrator positional accuracies are different for our two fields (1.53~mas in right ascension, 1.03~mas in declination for J1616+3621; 0.37~mas in right ascension, 0.56~mas in declination for J1623+3909; Petrov~\cite{petr07}), we give the following overall values. In Field A, the absolute positonal accuracy of the target sources is at least 1.6~mas in right ascension and 1.1~mas in declination. In Field B, these accuracies are 0.5~mas and 0.7~mas, respectively. Indeed, the comparison between the coordinates derived independently for the 3 target sources (J161748.4+380141, J161900.7+381139, J162111.2+374604; Table~\ref{fieldA}) located in the overlapping region of Fields~A and B (Fig.~\ref{skyplot}) shows that the diferences are within 2~mas in right ascension and within 1.7~mas in declination for each source. Due to the above reasons, for these common objects we consider the coordinates obtained in Field~B more accurate.   

The SDSS optical positions for all but 3 of the 24 VLBI-detected sources are within 180~mas from the peak positions measured in the VLBI images (Col.~10 in Table~\ref{fieldA}). This is in good agreement with what is expected and found for the SDSS offsets with respect to the VLBI positions for a large sample of common objects (e.g. Frey et al. \cite{frey06a,frey06b}). The extremely large ($\sim0\farcs7-0\farcs8$) VLBI--SDSS positional differences seen in the case of J161432.1+394445 and J162605.1+404806 might be due to a significant displacement between the radio and optical brightness peaks, most likely because of an arcsecond-scale extended radio emission. Our VLBI positions are in fact closer to (within $0\farcs5$ of) the FIRST radio positions which are accurate to $\sim1\arcsec$ only. 
Another possibility is that we see random associations. The chance of finding pairs of unrelated sources in our small 26-element sample that are projected so close to each other on the sky is not negligible: Ivezi\'c et al. (\cite{ivez02}) estimate 3\% probability for random matches between SDSS and FIRST objects within our search radius of $1\farcs5$. Therefore we choose a conservative approach and do {\it not} consider J161432.1+394445 and J162605.1+404806 successful VLBI detections of the corresponding SDSS quasars.

\subsection{Notes on individual sources}

Our two phase-reference calibrators (Figs.~\ref{a-cal}-\ref{b-cal}), as well as three of our brightest target sources (J161748.4+380141, J162240.7+381637 and J162330.5+355933; Figs.~\ref{images}g, q and l, respectively) are found in the VLBA Imaging and Polarimetry Survey (VIPS; Helmboldt et al. \cite{helm07})\footnote{\tt http://www.phys.unm.edu/$\sim$gbtaylor/VIPS} image data base. This latter survey concentrates on the celestial area covered by SDSS and provides polarization-sensitive 5-GHz imaging data obtained using the NRAO Very Long Baseline Array (VLBA). Their sample of more than 1100 sources was drawn from the VLA Cosmic Lens All-Sky Survey (CLASS; Myers et al. \cite{myer03}), selected on the basis of flat radio spectrum and a lower limit of 85~mJy flux density at 8.4~GHz. The brightness distributions seen in our images are consistent with the VIPS results for all the common sources.  

{\bf J161446.9+374607} 
The compact structure observed is consistent with the blazar classification (March\~a et al. \cite{marc01}). The 5-GHz flux density we detected is more than twice as high as the total flux density found in the GB6 catalogue. This source (Fig.~\ref{images}b) might therefore be highly variable. 

{\bf J162111.2+374604} 
Marecki et al. (\cite{mare06}) describe this quasar (known also as 4C~+37.46) as an asymmetric core-jet-lobe source. They imaged it with MERLIN at 5~GHz, and with VLBI at 1.7 and 5~GHz as part of a subarcsecond-size Compact Steep Spectrum (CSS) radio source sample. Our image in Fig.~\ref{images}k shows the inner jet structure extended to the north-east. Guided by the lower resolution images of Marecki et al. (\cite{mare06}), we were also able to marginally detect a small fraction ($\sim1$~mJy/beam) of the diffuse radio emission originating from the lobe at around 368~mas east and 376~mas north of the core (out of the field displayed in Fig.~\ref{images}k).  

{\bf J162229.3+400643}
The 5-GHz flux density we measured is significantly higher than the GB6 value, possibly indicating variability of the source (Fig.~\ref{images}p).
Neumann et al. (\cite{neum94}) give $S_{\rm{5~GHz}}=58$~mJy and a flat radio spectrum.
 
{\bf J162422.0+392440}
Two other compact FIRST radio sources are found within $13\arcsec$ of this object. These three form a nearly linear structure with an angular extent of $\sim21\arcsec$, the central object being the brightest. This is a double-lobed FR-II radio quasar (de Vries et al. \cite{devr06}) of which we imaged the core with VLBI (Fig.~\ref{images}r). It appears compact with a hint on an extension to the N-NE which might have been resolved with longer north--south baselines.

{\bf J162453.4+375806}
A high-redshift ($z=3.38$) quasar with a resolved mas-scale structure in the NE-SW direction (Fig.~\ref{images}s).
Benn et al. (\cite{benn05}) describe this source as the most radio-luminous broad absorption line (BAL) quasar known, with unusual radio and optical properties. 
Its radio emission is strongly polarised (11\% at 10~GHz) and the rotation measure (18350~rad~m$^{-2}$) is among the highest values known for any extragalactic source.
The GHz-peaked radio spectrum of the source suggests a young age.
Based on their 22-GHz VLA observations, Benn et al. (\cite{benn05}) give $0\farcs4\pm0\farcs1$ for the angular size of J162453.4+375806. On the other hand, the spectral turnover frequency ($\sim2$~GHz in the rest frame) indicates $\sim0.1-1$~kpc linear size.
The integrated flux density in our VLBI components (23.6~mJy) is practically equal to the GB6 value (22~mJy) and the total flux density measured by Benn et al. (\cite{benn05}) (23.3~mJy). Thus we can conclude that the entire radio emission comes from a region as small as $\sim10$~mas, corresponding to a projected linear size of less than 0.1~kpc (assuming a flat cosmological model with $\Omega_{m}=0.3$, $\Omega_{\Lambda}=0.7$ and $H_{\rm{0}}=70$~km~s$^{-1}$~Mpc$^{-1}$; this model is used throughout the paper).

{\bf J163020.7+375656}
We did not detect this object with VLBI at 5 GHz (peak brightness $<0.7$~mJy/beam). McLeod \& McLeod (\cite{mcle01}) studied the host galaxy of this optically luminous low-redshift ($z=0.39$) quasar which they call radio-quiet with the Near-Infrared Camera and Multi-Object Spectrometer (NICMOS) on the Hubble Space Telescope (HST). 

{\bf J163302.1+392427}
Another FIRST source is found within $9\arcsec$ to the south-east of our object. Its direction is close to the position angle of the mas-scale extension seen in our VLBI image (Fig.~\ref{images}v). Intra-night optical variability was detected in this quasar at one epoch by Stalin et al. (\cite{stal05}).


\section{Conclusions}
\label{sect6}

Phase-referenced 5-GHz VLBI images of 24 radio-loud AGNs at mJy flux density level or higher are presented here. To our knowledge, 20 of these sources have never been imaged with VLBI before. 
Five of our target sources (J161446.9+374607, J161748.4+380141, J162229.3+400643, J162330.5+355933, J162422.0+392440) are nearly as bright and compact at 5~GHz as our phase-reference calibrator sources. It is interesting to note that two of them (J161446.9+374607 and J162422.0+392440) have so far escaped from being selected for any of the VLBI surveys. This indicates that the radio sky is still not fully explored even for relatively strong ($\sim80$~mJy) mas-scale compact extragalactic sources. 

We detected 22 out of 26 target SDSS quasars, 85\% of our initial sample. (Two other VLBI detections are probably chance positional coincidences with SDSS quasars.)
Combining this with the DEVOS pilot project results (Paper~I) and counting also the phase-reference calibrator sources, the VLBI detection rate is nearly 90\% for the radio quasars that are known to be compact in both optical (SDSS) and in the FIRST survey, and have integrated flux density $S_{\rm{1.4}}>20$~mJy in FIRST.
Based on currently available optical and low-resolution radio sky surveys, we found a highly efficient way to pre-select extragalactic sources that are detectable with VLBI at 5~GHz. On the other hand, for any given object found in the SDSS and FIRST lists, the selection method we used is also an excellent tool for assessing whether it can be potentially successfully imaged with VLBI.

The SDSS DR6 contains nearly 9000 quasars that satisfy our selection criteria and thus can be targeted with VLBI. Extrapolating our detection rate for this entire sample, the number of potential VLBI detections can be as high as 7500. Such a big VLBI-optical sample --~partly with spectroscopic redshifts available~-- would eventually be invaluable for a multitude of astrophysical and cosmological studies. 

Astrometric projects like global or local densifications of the ICRF, and linking the radio reference frame with the future optical reference frame would also benefit from the DEVOS data. In practice, the data base is to be built up field by field gradually, concentrating on celestial areas that are otherwise targeted by different projects, e.g. deep spectroscopy, or deep imaging at any waveband. The immediate science return of these observations could be enhanced this way.     


\begin{acknowledgements}
We thank the anonymous referee for suggestions that helped improving the discussion of our results.
The EVN is a joint facility of European, Chinese, South African and other radio astronomy institutes funded by their national research councils.
This work has benefited from research funding from the European Community's sixth Framework Programme under RadioNet R113CT 2003 5058187 and from the Hungarian Scientific Research Fund (OTKA, grant
no.\ T046097). 

This research has made use of the NASA/IPAC Extragalactic Database (NED) which
is operated by the Jet Propulsion Laboratory, California Institute of
Technology, under contract with the National Aeronautics and Space
Administration (NASA). This research has made use of NASA's Astrophysics Data 
System.
Funding for the Sloan Digital Sky Survey (SDSS) and SDSS-II has been provided by the Alfred P. Sloan Foundation, the Participating Institutions, the National Science Foundation, the U.S. Department of Energy, the National Aeronautics and Space Administration, the Japanese Monbukagakusho, and the Max Planck Society, and the Higher Education Funding Council for England. The SDSS Web site is http://www.sdss.org/.
The SDSS is managed by the Astrophysical Research Consortium (ARC) for the Participating Institutions. 

\end{acknowledgements}



\Online

\begin{figure*}
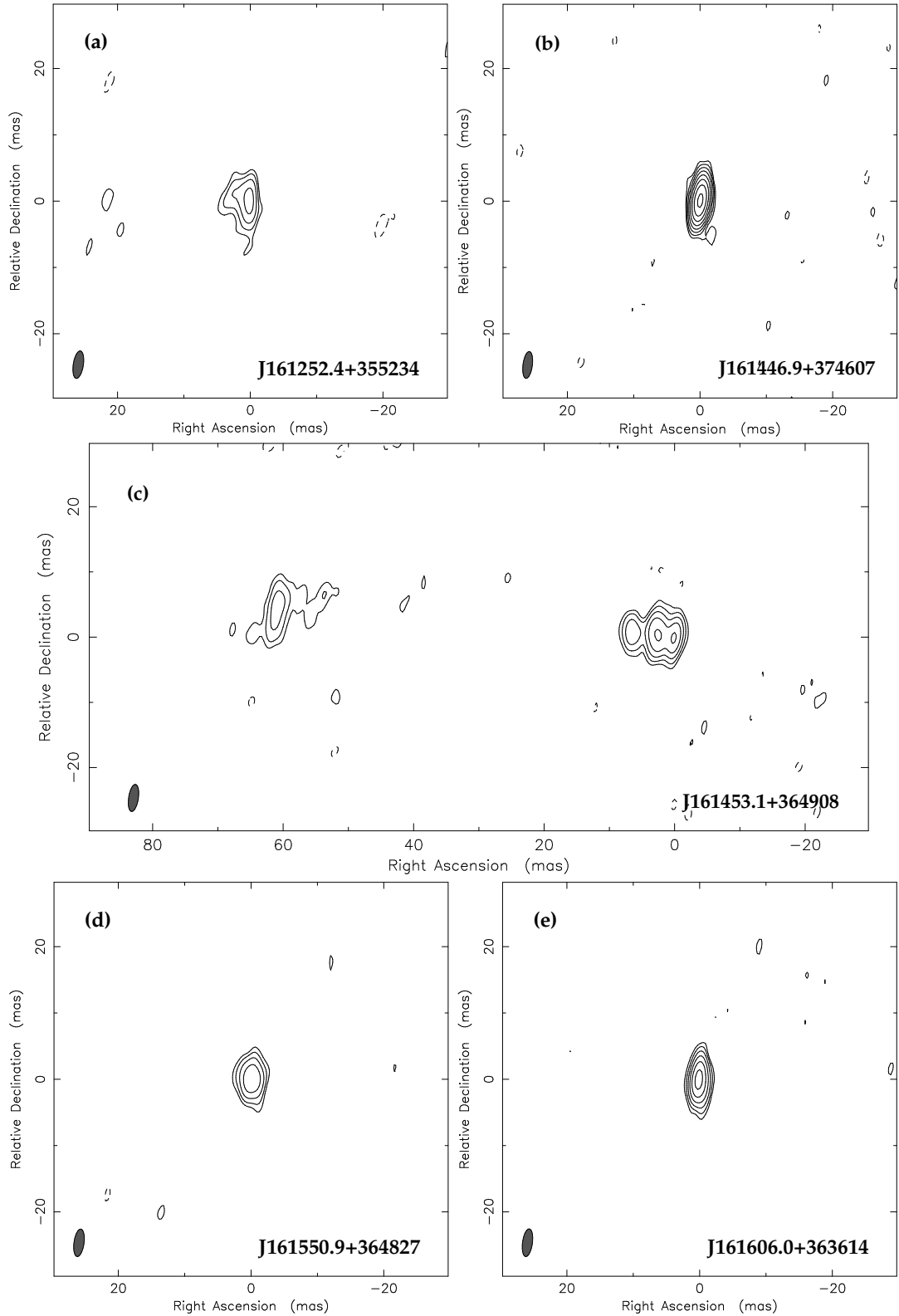

\setcounter{figure}{4}
  \centering
    \includegraphics[width=7cm,bb= 70 169 522 626,angle=270,clip= ]{8711-f5a.ps}
    \includegraphics[width=7cm,bb= 70 177 506 619,angle=270,clip= ]{8711-f5b.ps}
    \includegraphics[width=7cm,bb= 102 27 490 768,angle=270,clip= ]{8711-f5c.ps} 
    \includegraphics[width=7cm,bb= 70 169 522 626,angle=270,clip= ]{8711-f5d.ps}
    \includegraphics[width=7cm,bb= 70 169 522 626,angle=270,clip= ]{8711-f5e.ps}
  \caption{Naturally weighted 5-GHz VLBI images of 24 programme sources. The
image parameters (peak brightness, lowest contour level corresponding to $3\sigma$ image noise,
Gaussian restoring beam size and orientation) are listed and explained in 
Table~\ref{image-par}. 
The positive contour levels increase by a factor of 2. The coordinates are 
relative to the bightness peak for which the absolute coordinates are 
given in Table~\ref{fieldA}. The restoring beam is indicated with 
an ellipse in the lower-left corner.}
  \label{images}
\end{figure*}

\addtocounter{figure}{-1}
\begin{figure*}
  \centering
    \includegraphics[width=7cm,bb= 70 169 522 626,angle=270,clip= ]{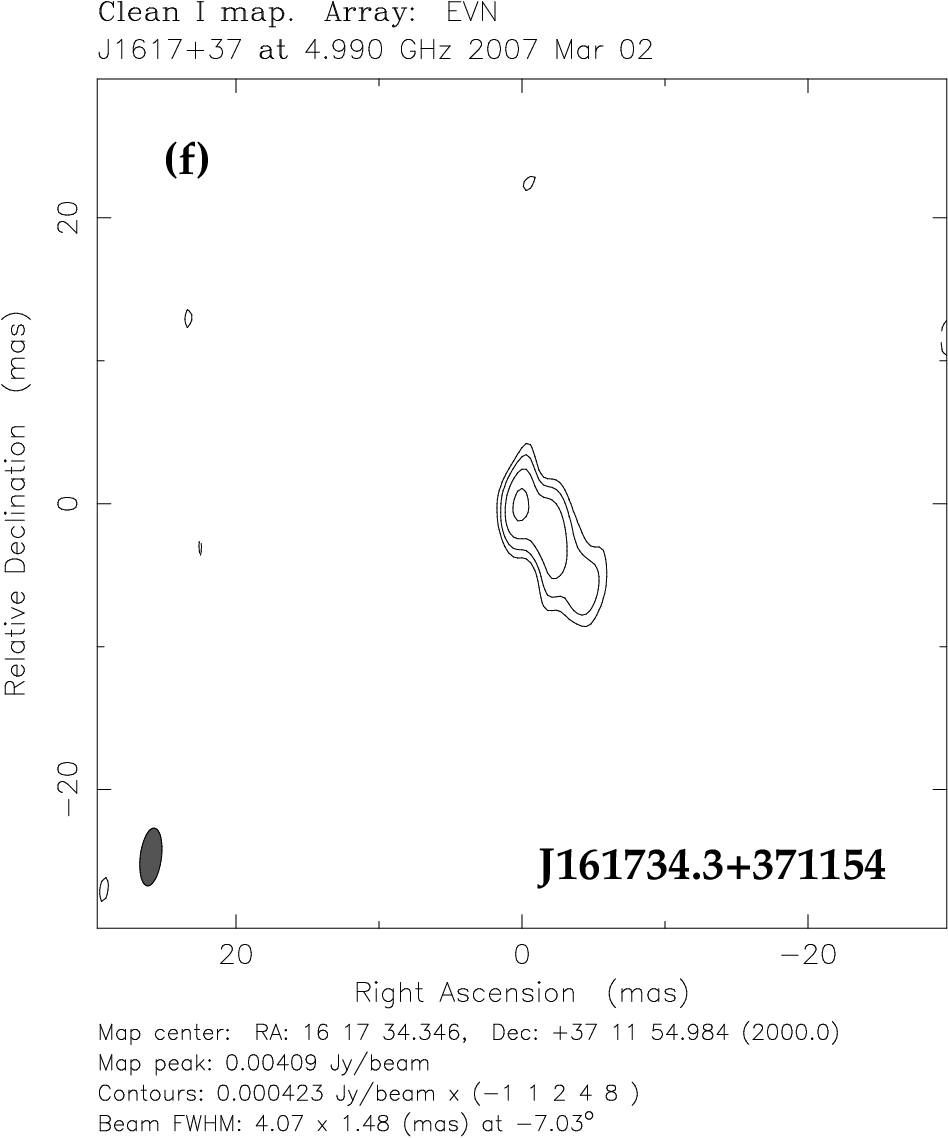}
    \includegraphics[width=7cm,bb= 70 177 506 619,angle=270,clip= ]{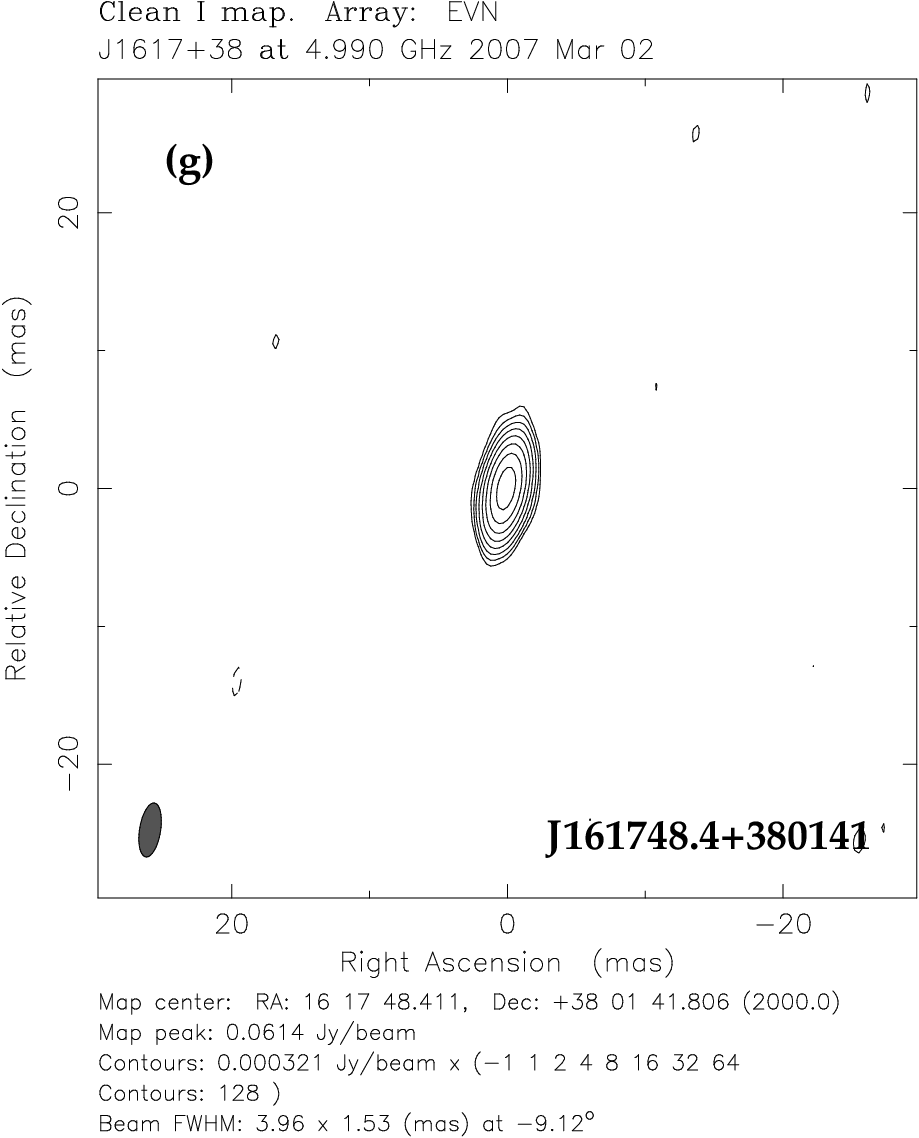}
    \includegraphics[width=7cm,bb= 70 169 522 626,angle=270,clip= ]{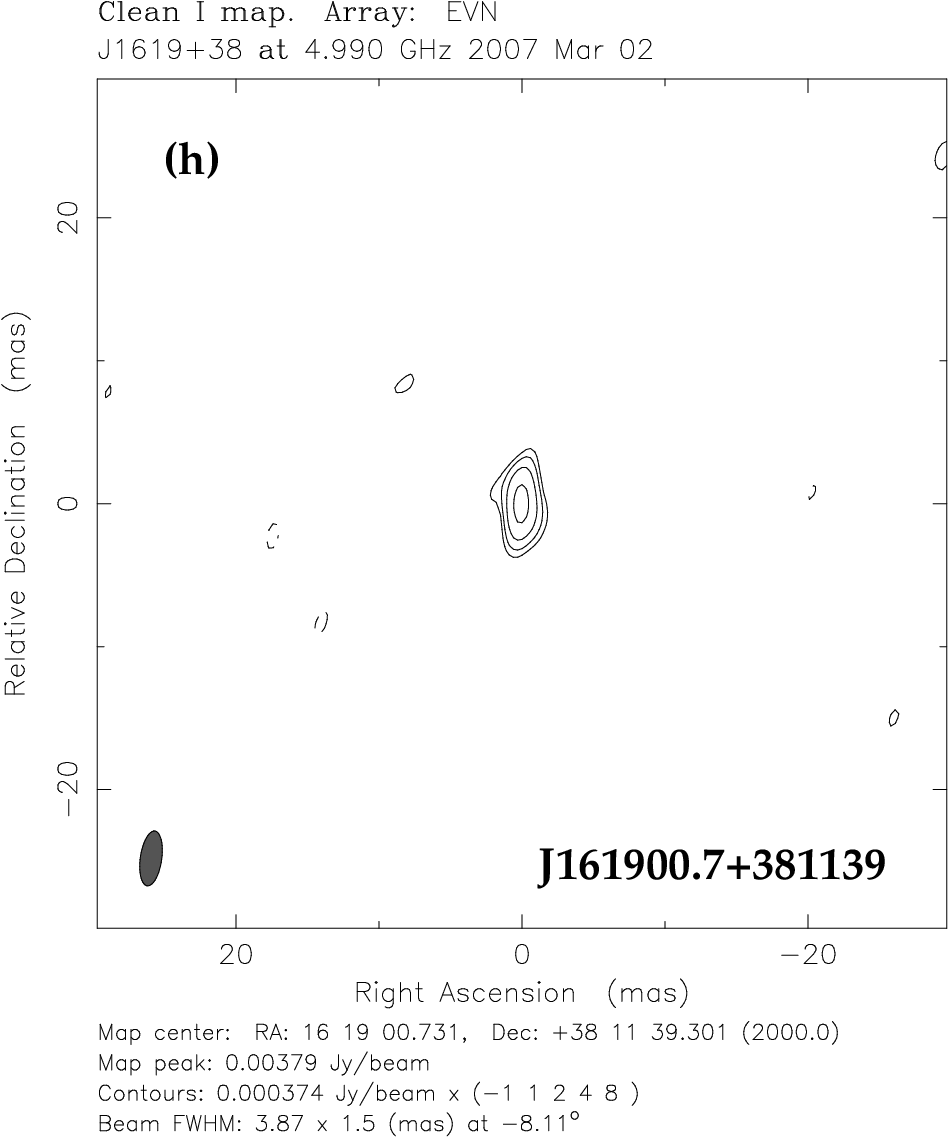}
    \includegraphics[width=7cm,bb= 70 169 522 626,angle=270,clip= ]{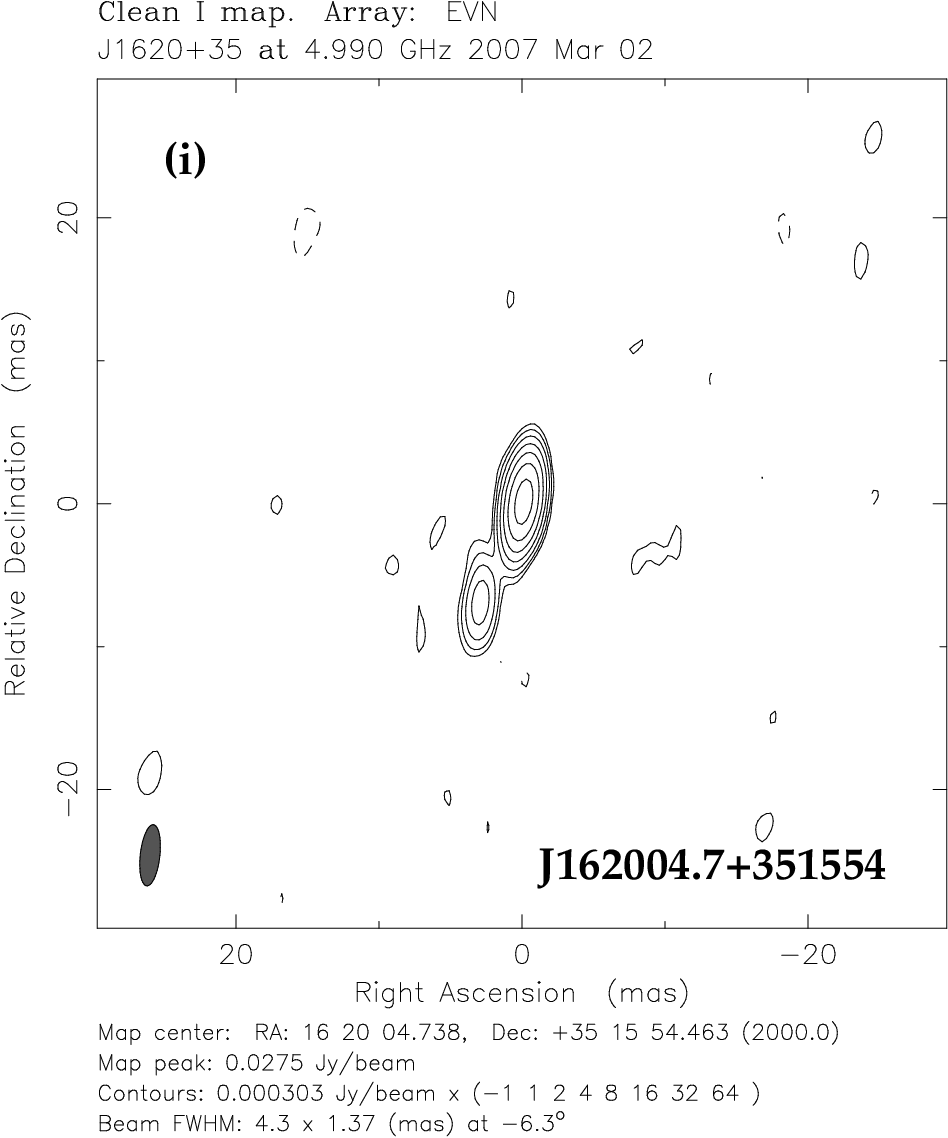}
    \includegraphics[width=7cm,bb= 70 169 522 626,angle=270,clip= ]{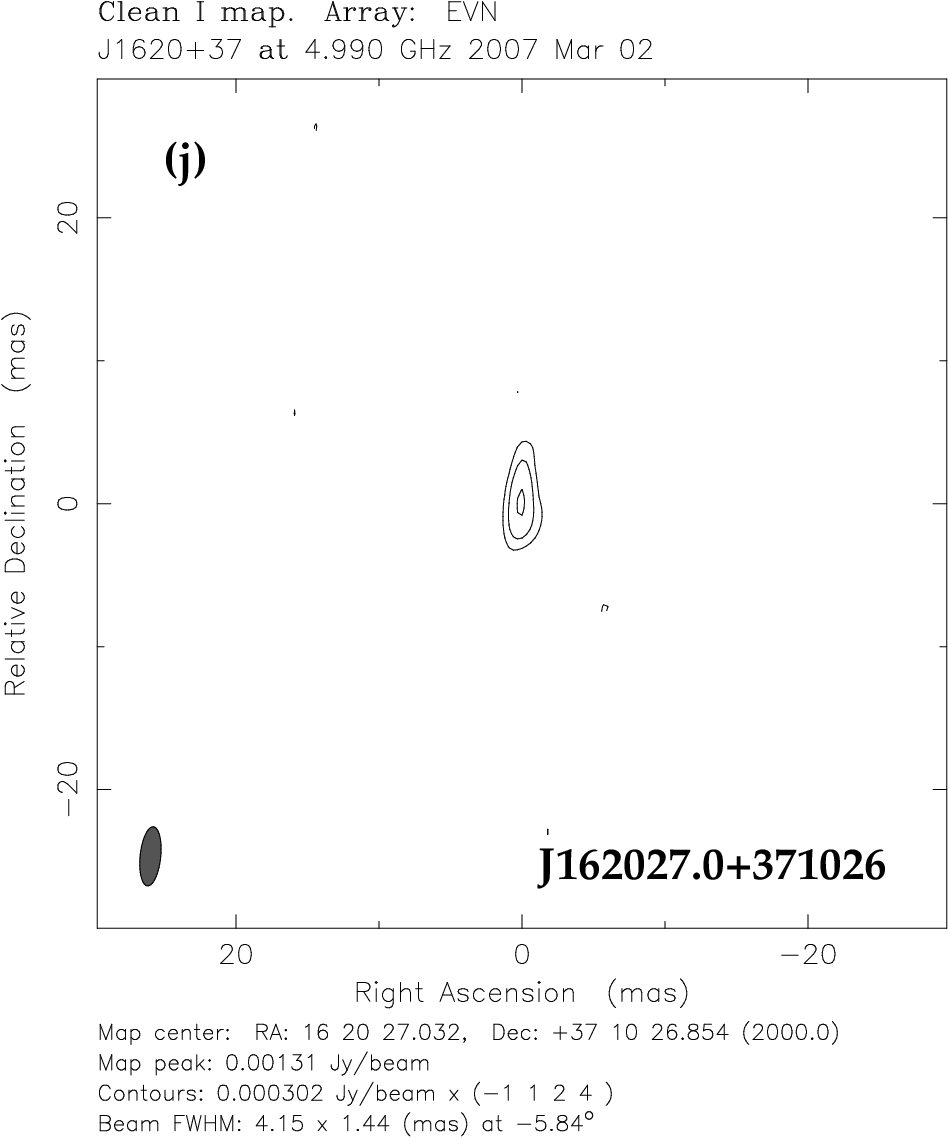}
    \includegraphics[width=7cm,bb= 70 169 522 626,angle=270,clip= ]{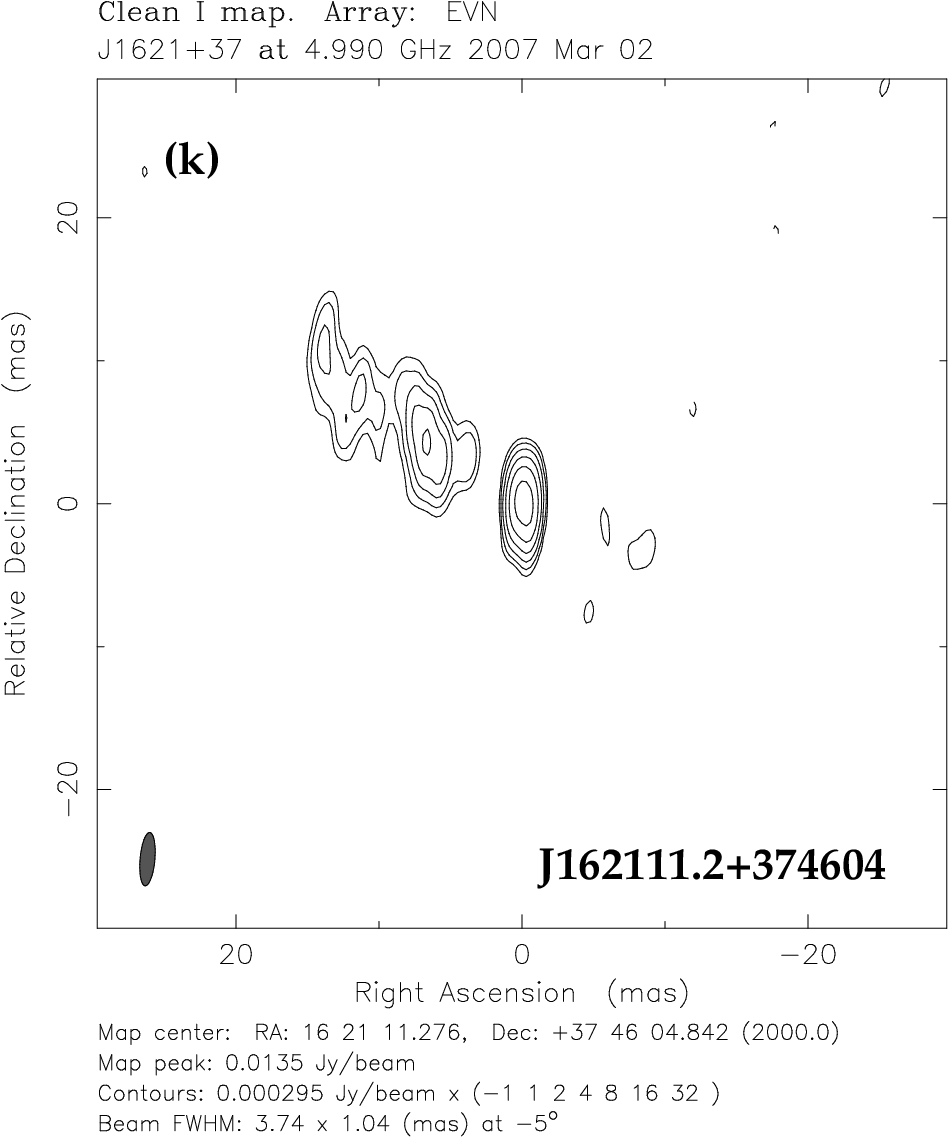}
  \caption{{\it (continued)}}
\end{figure*}

\addtocounter{figure}{-1}
\begin{figure*}
  \centering
    \includegraphics[width=7cm,bb= 70 169 522 626,angle=270,clip= ]{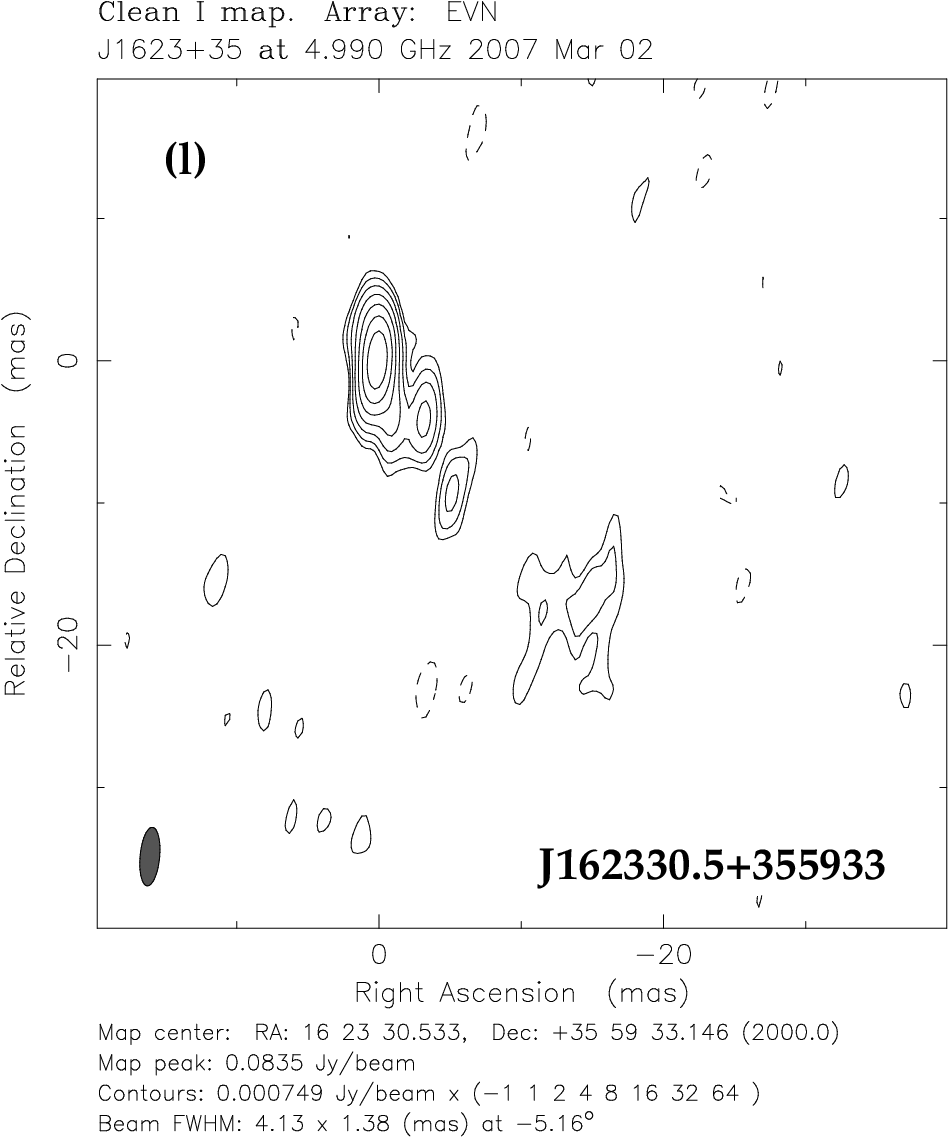}
    \includegraphics[width=7cm,bb= 70 169 522 626,angle=270,clip= ]{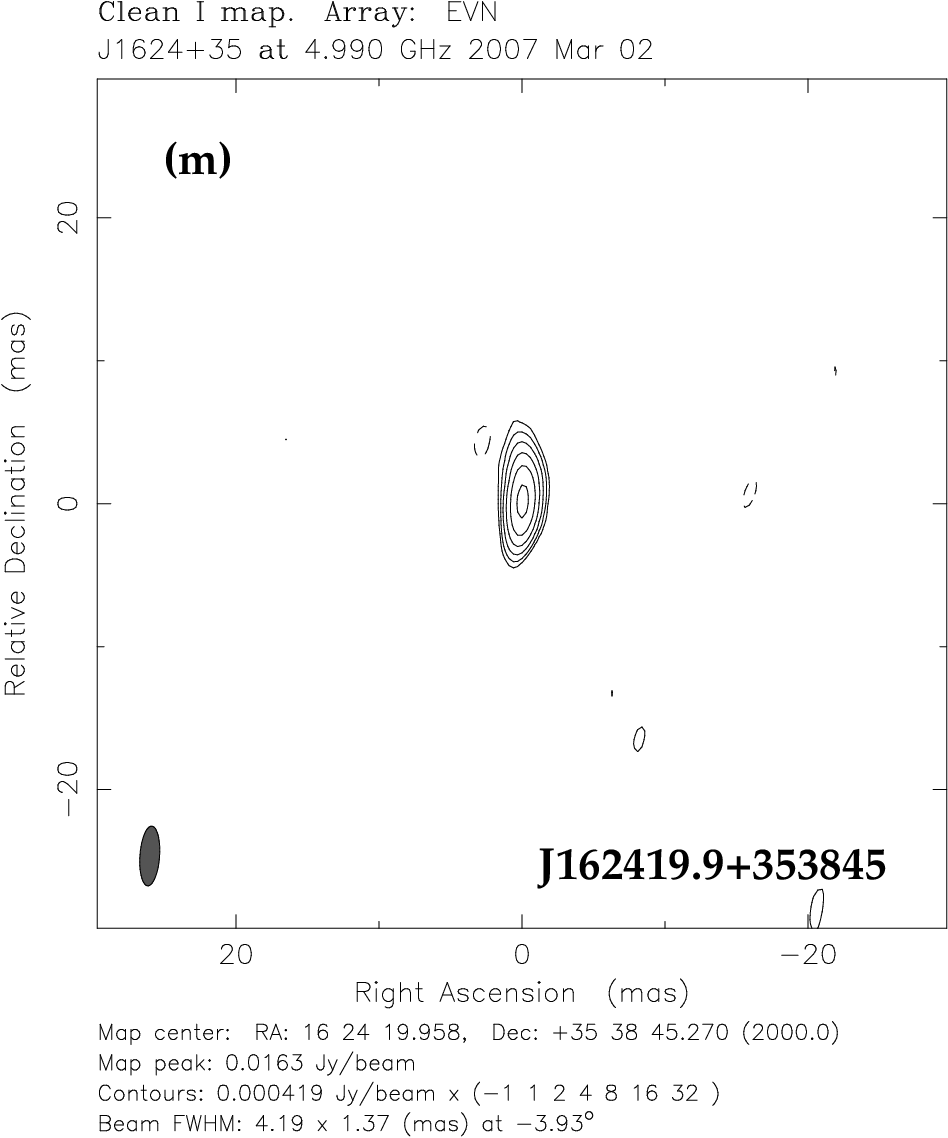}
    \includegraphics[width=7cm,bb= 70 169 522 626,angle=270,clip= ]{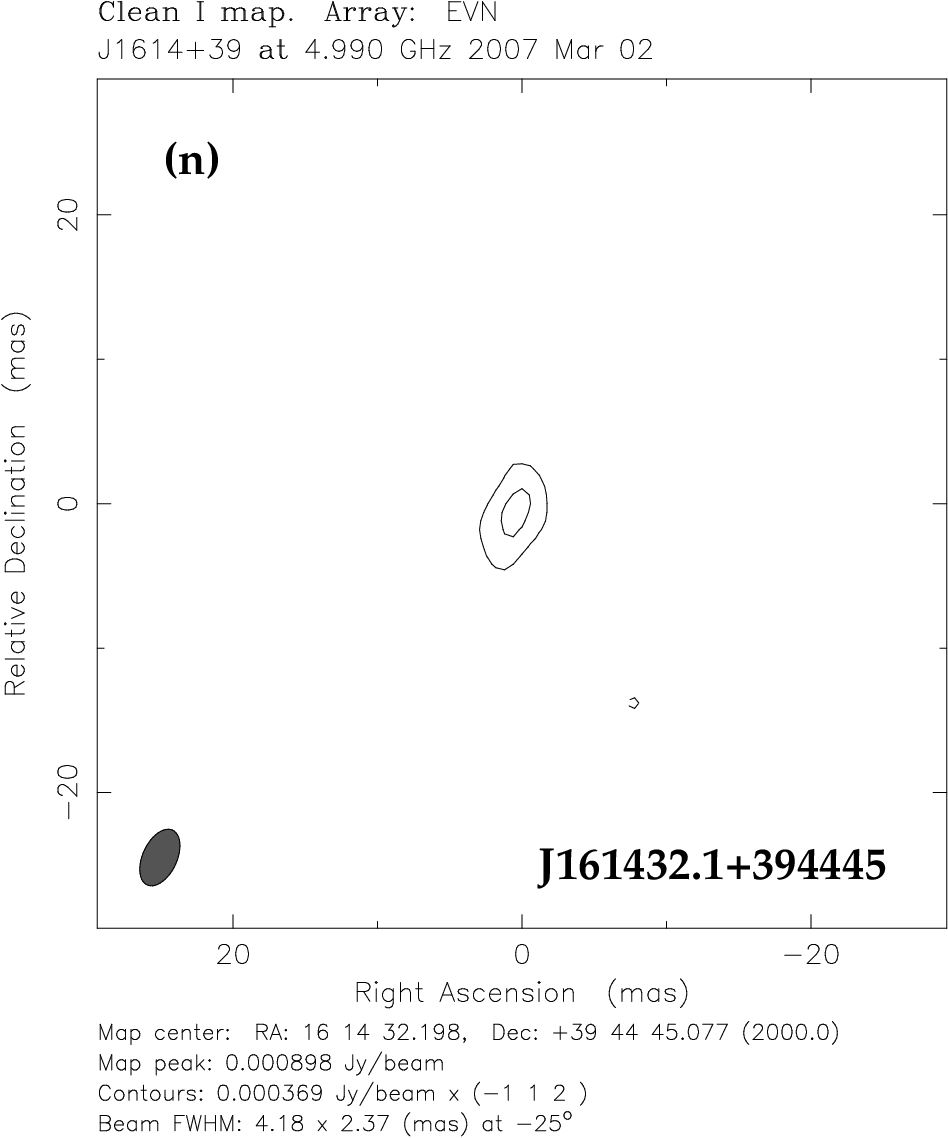}
    \includegraphics[width=7cm,bb= 70 169 522 626,angle=270,clip= ]{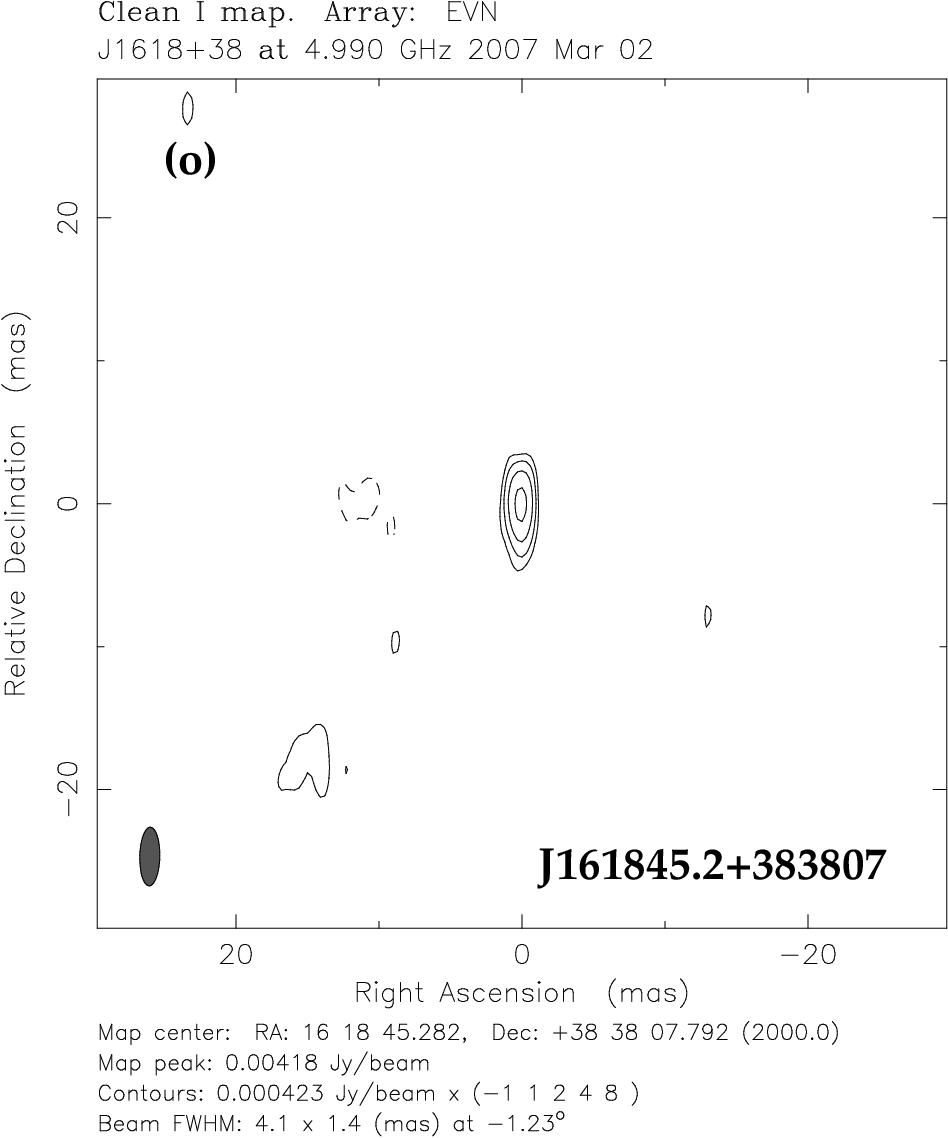}
    \includegraphics[width=7cm,bb= 70 177 506 619,angle=270,clip= ]{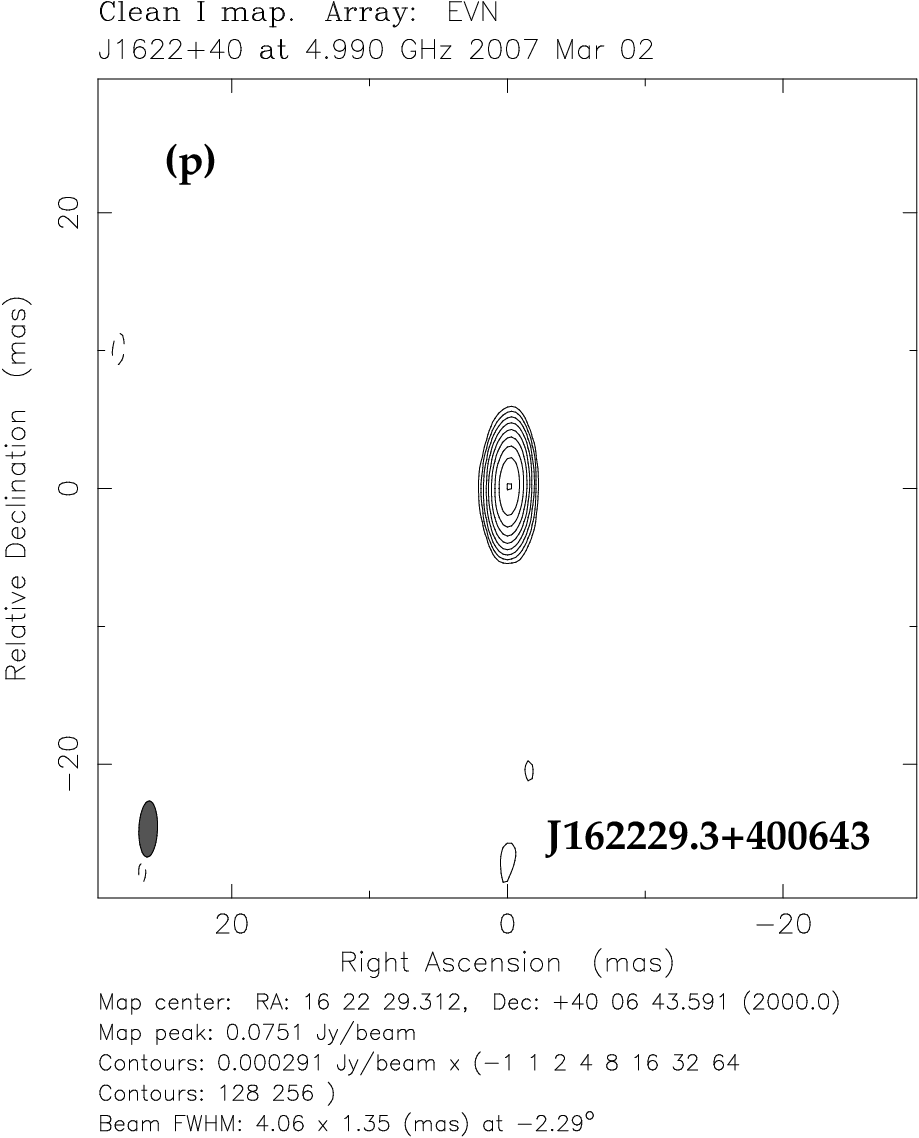}
    \includegraphics[width=7cm,bb= 70 177 506 619,angle=270,clip= ]{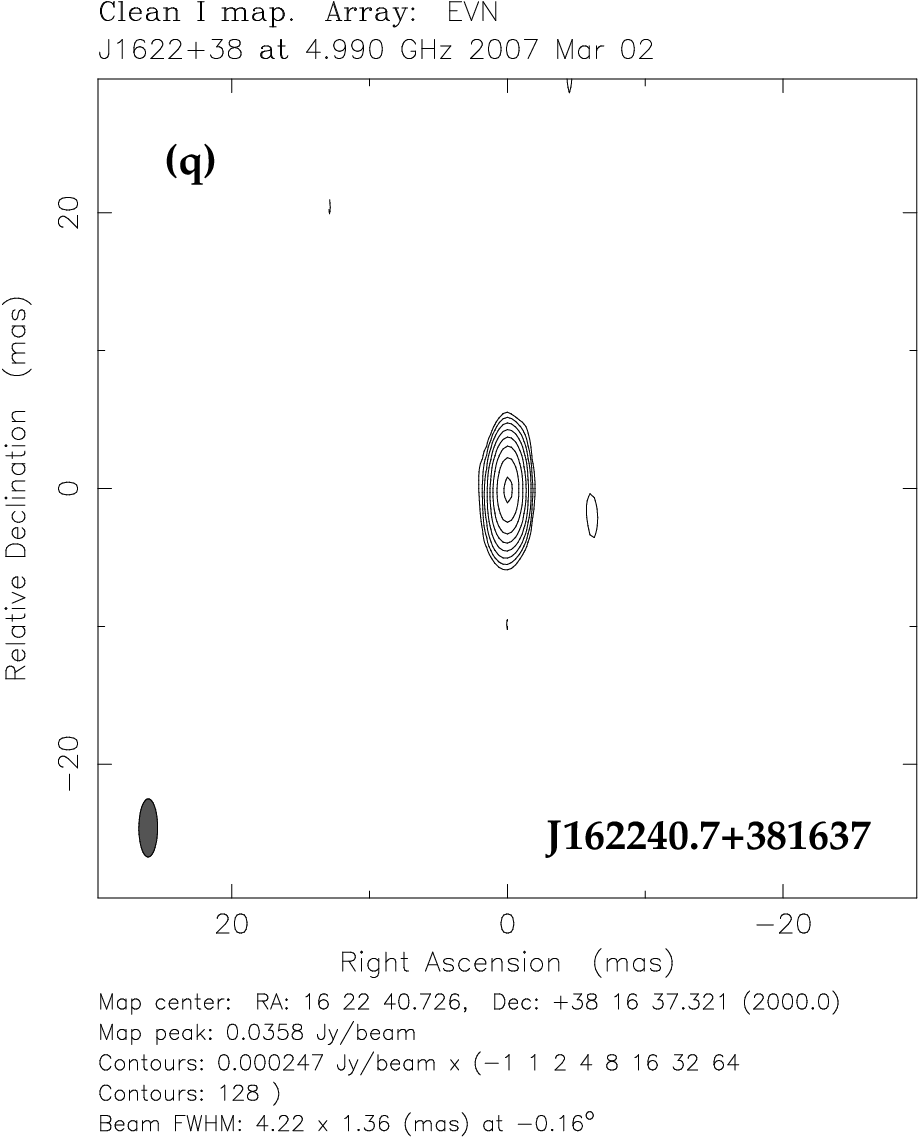}
  \caption{{\it (continued)}}
\end{figure*}

\addtocounter{figure}{-1}
\begin{figure*}
  \centering
    \includegraphics[width=7cm,bb= 70 177 506 619,angle=270,clip= ]{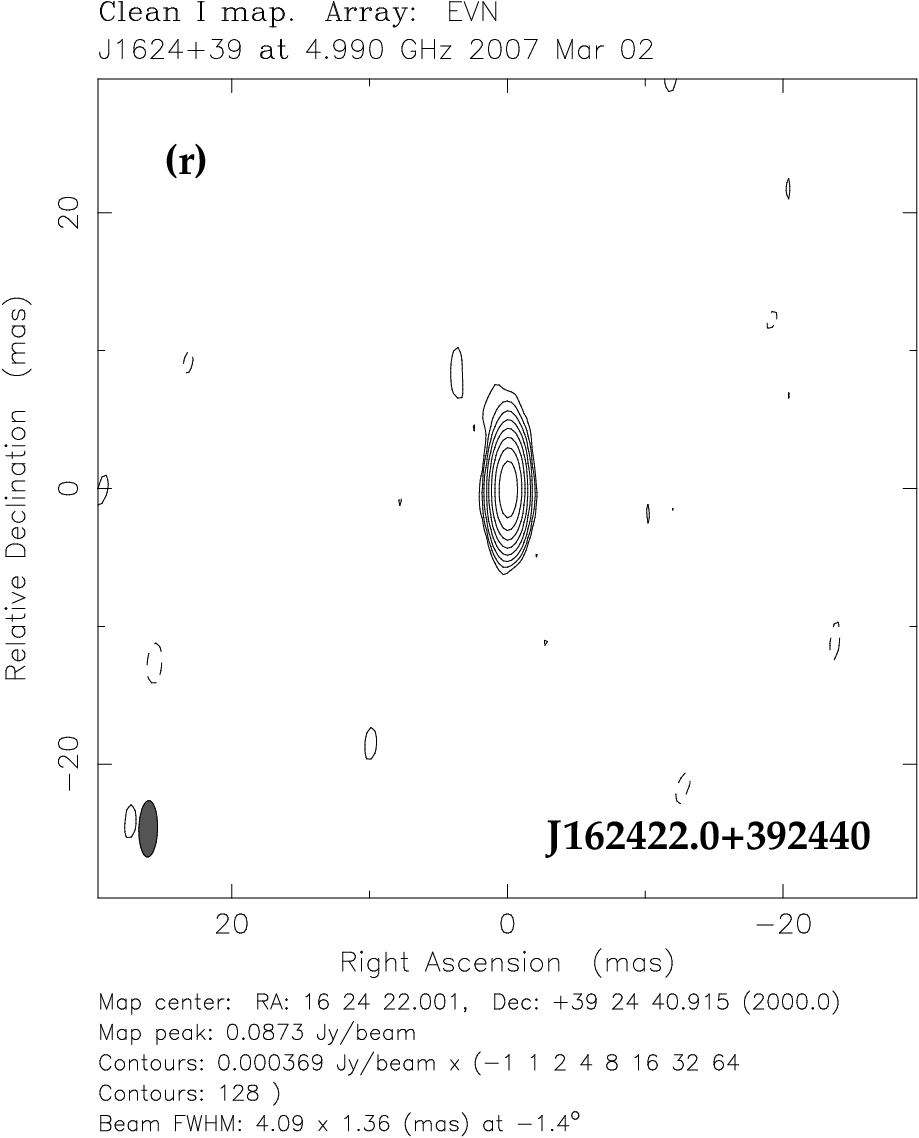}
    \includegraphics[width=7cm,bb= 70 169 522 626,angle=270,clip= ]{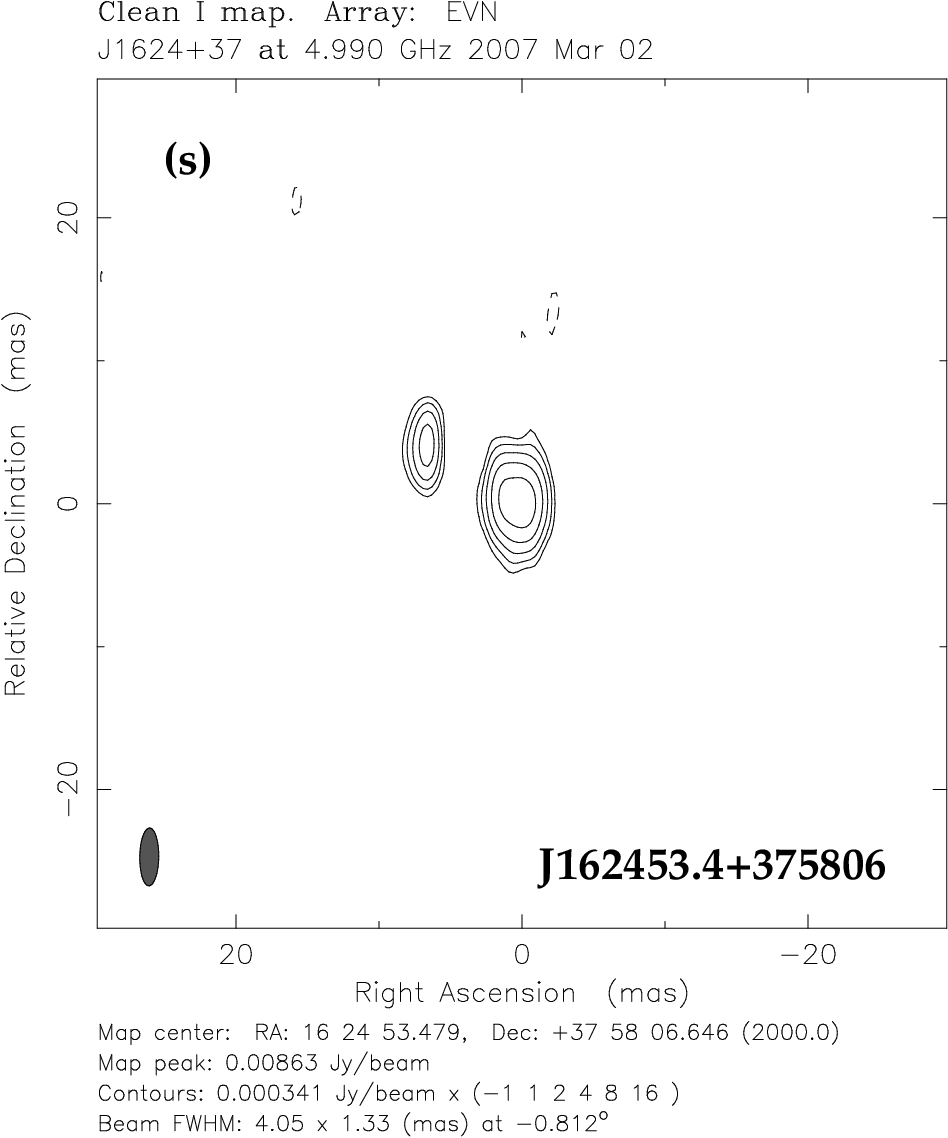}
    \includegraphics[width=7cm,bb= 70 169 522 626,angle=270,clip= ]{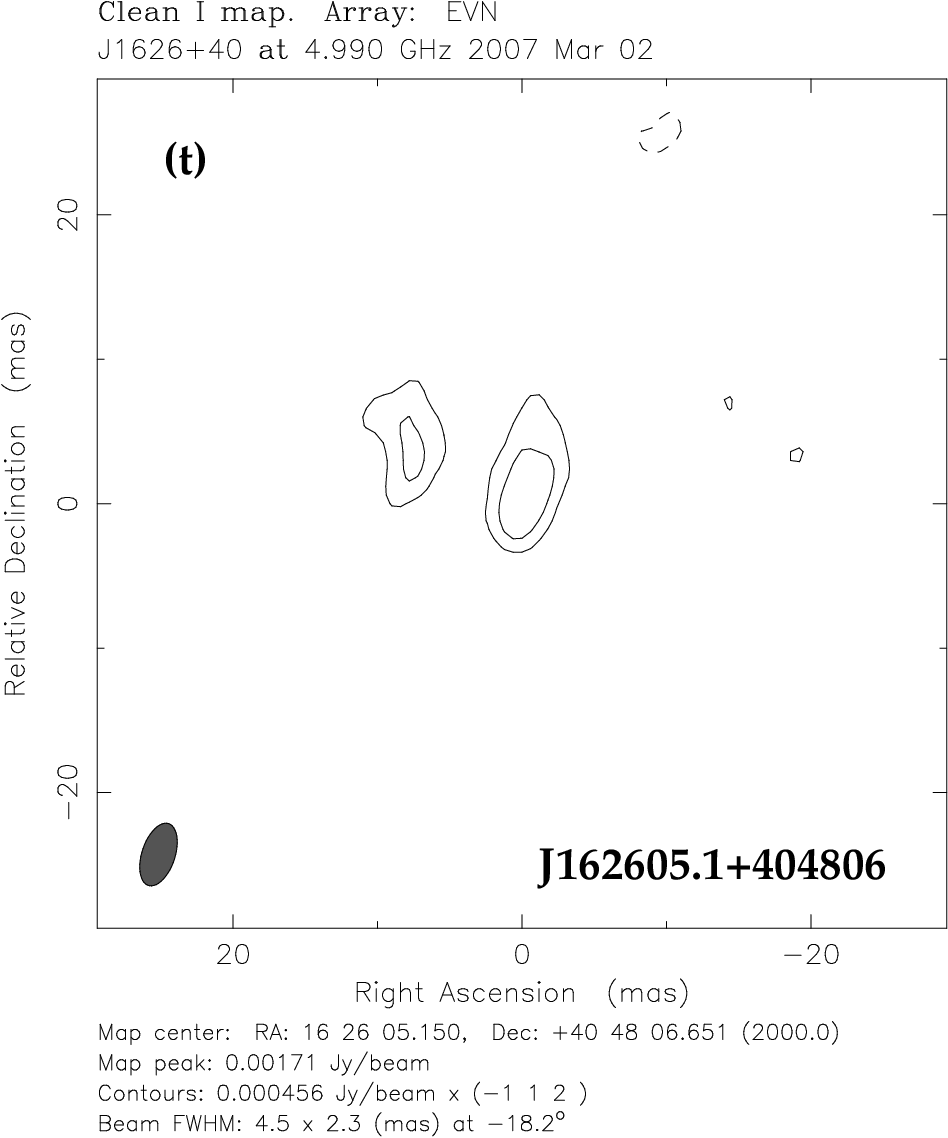}
    \includegraphics[width=7cm,bb= 70 169 522 626,angle=270,clip= ]{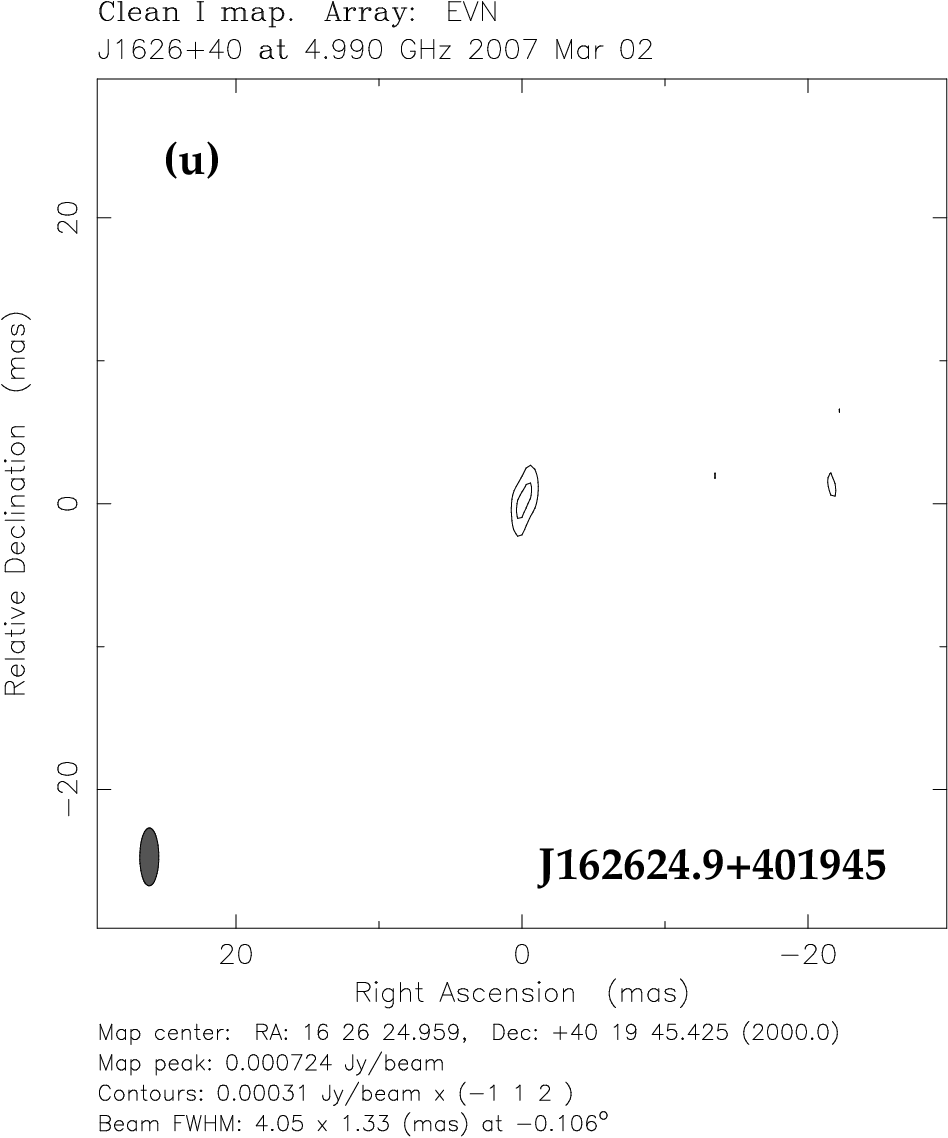}
    \includegraphics[width=7cm,bb= 70 169 522 626,angle=270,clip= ]{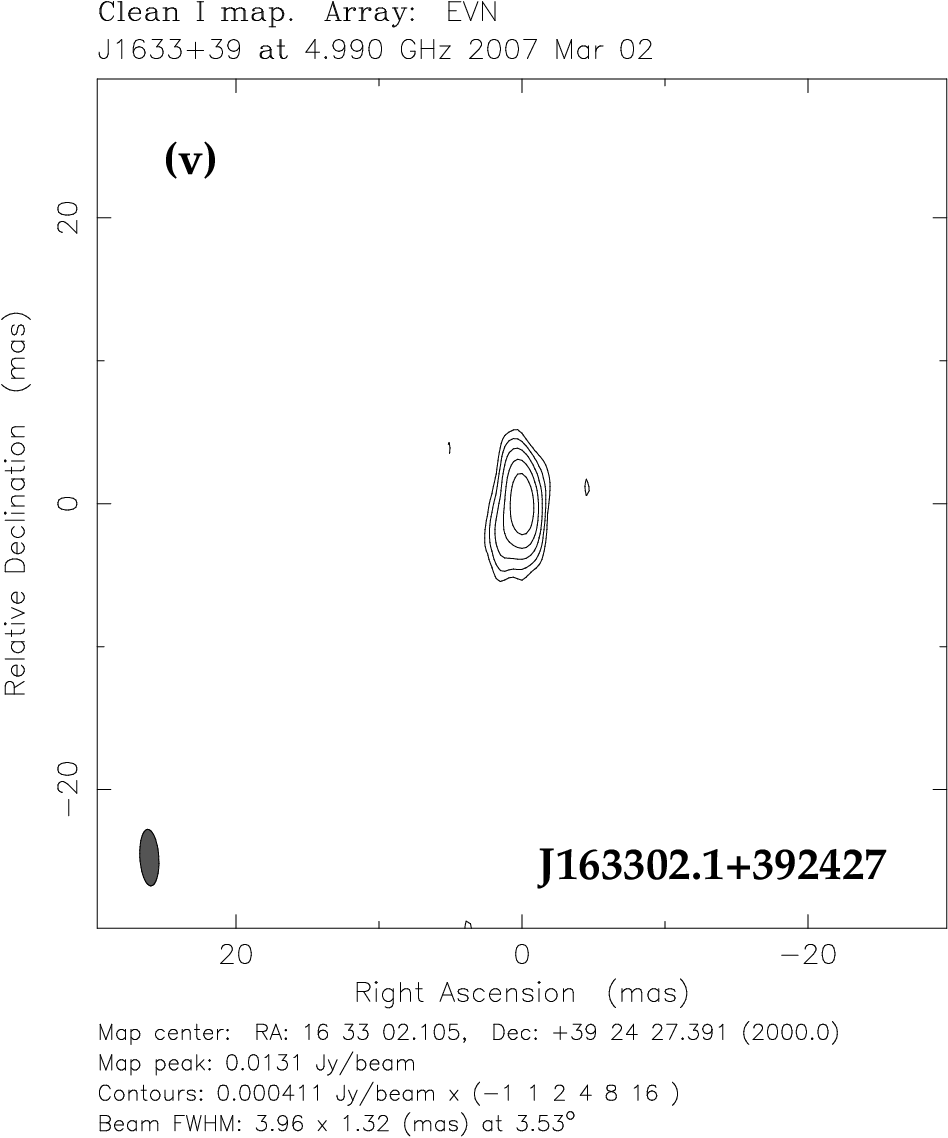}
  \caption{{\it (continued)}}
\end{figure*}

\addtocounter{figure}{-1}
\begin{figure*}
  \centering
    \includegraphics[width=14cm,bb= 72 271 523 523,angle=270,clip= ]{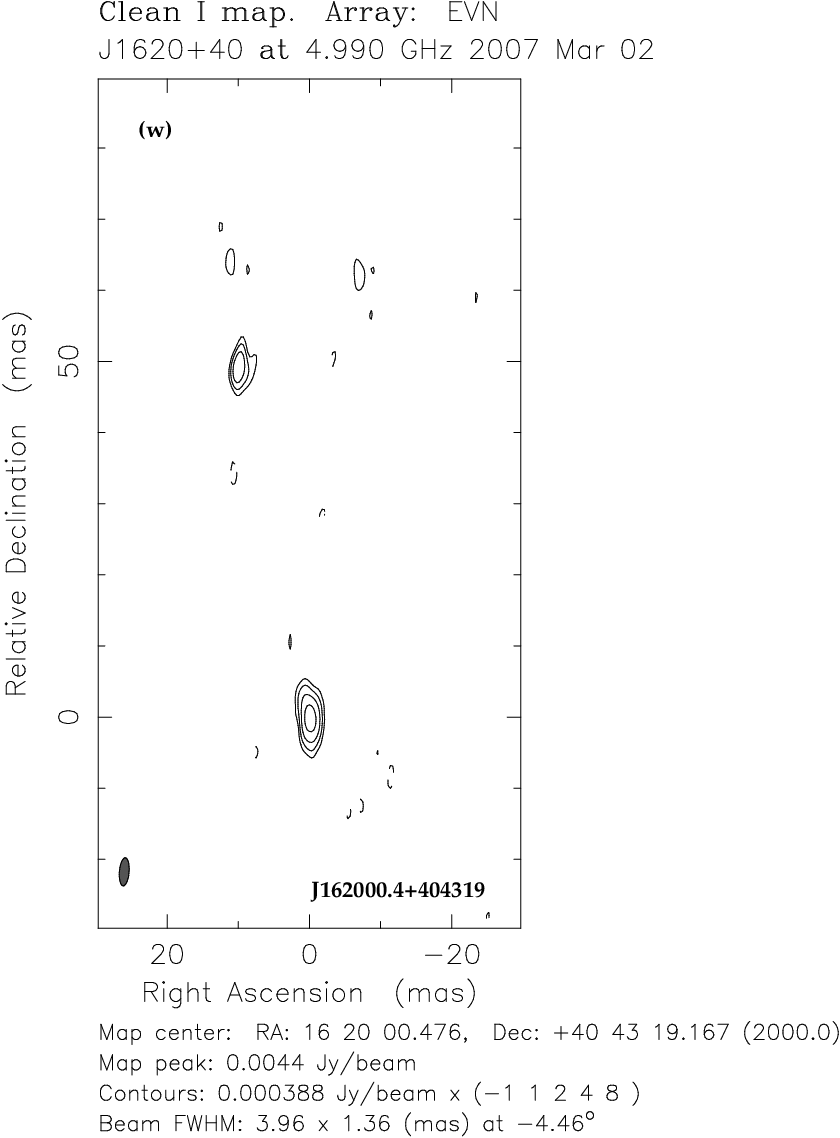} 
    \includegraphics[width=7cm,bb= 70 66 522 729,angle=270,clip= ]{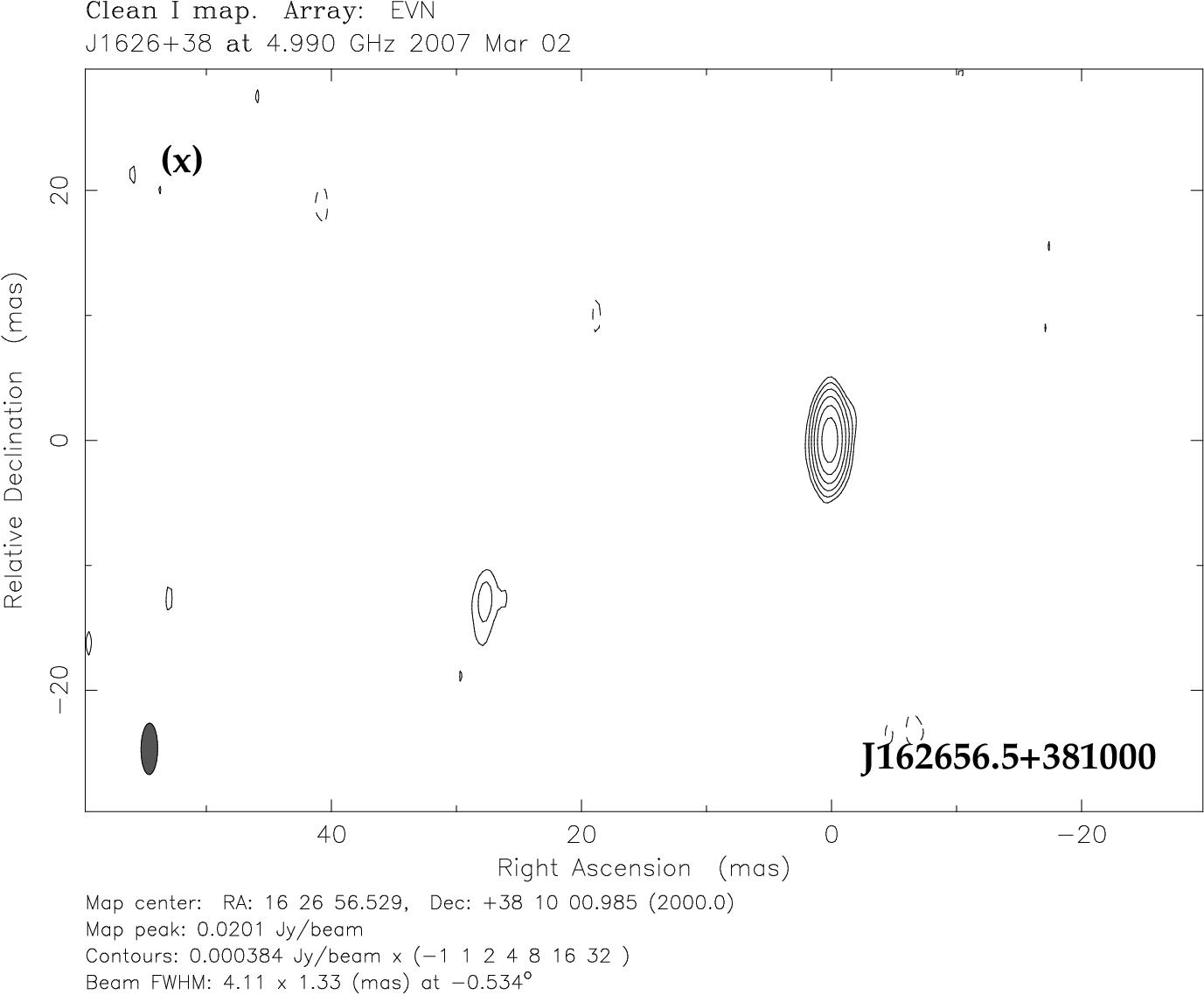} 
  \caption{{\it (continued)}}
\end{figure*}

\begin{table}
 \caption{VLBI image parameters for Field A ({\it top}) and Field B ({\it bottom})}
 \label{image-par}
 \begin{tabular}{@{}lrclr}
  \hline\hline
  Source & Peak       & Lowest  & \multicolumn{2}{c}{Restoring beam} \\
  name   &            & contour & size  & PA  \\
         & \multicolumn{2}{c}{(mJy/beam)} & (mas$\times$mas) & ($\degr$) \\
 \hline
J161252.4+355234  &  3.7 & $\pm$0.33 & $ 4.2\times 1.5$ & $-8.6$ \\
J161446.9+374607  & 76.9 & $\pm$0.50 & $ 4.1\times 1.4$ & $-7.4$  \\
J161453.1+364908  &  6.8 & $\pm$0.39 & $ 4.2\times 1.5$ & $-8.5$  \\
J161550.9+364827  &  4.8 & $\pm$0.31 & $ 4.2\times 1.5$ & $-7.7$  \\
J161606.0+363614  & 14.0 & $\pm$0.33 & $ 4.2\times 1.5$ & $-7.6$ \\
J161734.3+371154  &  4.1 & $\pm$0.42 & $ 4.1\times 1.5$ & $-7.0$  \\
J161748.4+380141  & 61.4 & $\pm$0.32 & $ 4.0\times 1.5$ & $-9.0$  \\
J161900.7+381139  &  3.8 & $\pm$0.37 & $ 3.9\times 1.5$ & $-8.1$  \\
J162004.7+351554  & 27.5 & $\pm$0.30 & $ 4.3\times 1.4$ & $-6.3$  \\
J162027.0+371026  &  1.3 & $\pm$0.30 & $ 4.2\times 1.4$ & $-5.8$  \\
J162111.2+374604  & 13.5 & $\pm$0.30 & $ 3.7\times 1.0$ & $-5.0$  \\
J162330.5+355933  & 83.5 & $\pm$0.75 & $ 4.1\times 1.4$ & $-5.2$  \\
J162419.9+353845  & 16.3 & $\pm$0.42 & $ 4.2\times 1.4$ & $-3.9$  \\
  \hline
J161432.1+394445  &  0.9 & $\pm$0.37 & $ 4.2\times 2.4$ & $-25.0$  \\
J161845.2+383807  &  4.2 & $\pm$0.42 & $ 4.1\times 1.4$ & $-1.2$  \\ 
J162000.4+404319  &  4.4 & $\pm$0.39 & $ 4.0\times 1.4$ & $-4.5$  \\ 
J162229.3+400643  & 75.1 & $\pm$0.29 & $ 4.1\times 1.4$ & $-2.3$  \\
J162240.7+381637  & 35.8 & $\pm$0.25 & $ 4.2\times 1.4$ & $-0.2$  \\ 
J162422.0+392440  & 87.3 & $\pm$0.37 & $ 4.1\times 1.4$ & $-1.4$  \\ 
J162453.4+375806  &  8.6 & $\pm$0.34 & $ 4.1\times 1.3$ & $-0.8$  \\
J162605.1+404806  &  1.7 & $\pm$0.46 & $ 4.5\times 2.3$ & $-18.5$  \\
J162624.9+401945  &  0.7 & $\pm$0.31 & $ 4.1\times 1.3$ & $-0.1$  \\
J162656.5+381000  & 20.1 & $\pm$0.38 & $ 4.1\times 1.3$ & $-0.5$  \\ 
J163302.1+392427  & 13.1 & $\pm$0.41 & $ 4.0\times 1.3$ & $3.5$  \\ 
  \hline
\end{tabular}
\\
Notes: Col.~1 -- source name; 
Col.~2 -- peak brightness at 5~GHz (mJy/beam); 
Col.~3 -- lowest contour level (mJy/beam) corresponding to $3\sigma$ image noise; 
Col.~4 -- Gaussian restoring beam size (mas$\times$mas);
Col.~5 -- restoring beam major axis position angle ($\degr$) measured from north through east.
\end{table}
\end{document}